\documentclass[12pt]{article}
\usepackage{jheppub}
\usepackage{a4wide,epsfig,psfrag,amsmath,amssymb,scalefnt}
\usepackage[dvipsnames]{xcolor}
\usepackage{braket}
\usepackage{hyperref}
\usepackage{subfig}
\usepackage{cleveref}
\usepackage{slashed}
\usepackage{booktabs}
\usepackage[separate-uncertainty = true]{siunitx}
\usepackage{placeins}

\graphicspath{ {figs/} }

\newcommand{\lt}{\left}
\newcommand{\rt}{\right}

\newcommand{\nn}{\nonumber\\}
\newcommand{\ov}{\overline}
\newcommand{\eq}[1]{Eq.~(\ref{#1})}
\newcommand{\eqsand}[2]{Eqs.~(\ref{#1}) and (\ref{#2})}

\newcommand{\gev}{\,\mbox{GeV}}

\newcommand{\imag}{\mbox{Im\,}}
\newcommand{\real}{\mbox{Re\,}}
\newcommand{\Bbar}{\bar{B}}

\newcommand{\bb}{\ensuremath{B\!-\!\Bbar\,}}
\newcommand{\bbs}{\ensuremath{B_s\!-\!\Bbar{}_s\,}}
\newcommand{\bbq}{\ensuremath{B_q\!-\!\Bbar{}_q\,}}
\newcommand{\bbm}{\bb\ mixing}
\newcommand{\bbms}{\bbs\ mixing}

\newcommand{\bbmq}{\bbq\ mixing}

\newcommand{\fig}[1]{Fig.~\ref{#1}}

\newcommand{\lqcd}{\Lambda_{\rm QCD}} 
\newcommand{\dm}{\ensuremath{\Delta M}}
\newcommand{\dg}{\ensuremath{\Delta \Gamma}}
\newcommand{\ds}{\displaystyle}

\newcommand{\gqbtf}{\ensuremath{\Gamma (\Bbar{}_q(t) \rightarrow f )}}

\newcommand{\gdtfb}{\ensuremath{\Gamma (B_q(t) \rightarrow \ov{f} )}}

\newcommand{\mev}{\mbox{MeV}}

\newcommand\numberthis{\addtocounter{equation}{1}\tag{\theequation}}

\usepackage{changes}

\allowdisplaybreaks

\abstract{
We compute next-to-next-to-leading order corrections to the decay
width difference of mass eigenstates and the charge-parity (CP) asymmetry
$a_{\rm fs}$ in flavour-specific decays of neutral $B$ mesons.  We
include both current-current and penguin operators at three-loop
order. All input integrals in the transition amplitude are reduced to
a small set of master integrals which depend on the ratio of the charm
and bottom quark masses.  The latter are computed using semi-analytic
methods which provide deep expansions around properly selected values
of $m_c/m_b$. We provide numerical results for $\Delta\Gamma$ and
$\Delta\Gamma/\Delta M$, both for the $B_d$ and $B_s$ system, including a
detailed uncertainty analysis.  Using the experimental value for the mass difference $\dm_s$ we predict $\dg_s=(0.078 \pm 0.015) ~\mbox{ps}^{-1}$. 
For the CP asymmetries we find $ a_{\rm fs}^s = (2.27 \pm 0.13 ) \times 10^{-5}$ and 
    $a_{\rm fs}^d = -(5.19 \pm 0.30)~\times~10^{-4}$.
Furthermore, we show that the ratios
$(\Delta\Gamma_s/\Delta M_s) / (\Delta\Gamma_d/\Delta M_d)$ 
and $\Delta\Gamma_d / \Delta\Gamma_s$ can be predicted with high precision. The former 
quantity permits the prediction \mbox{$\dg_d=(0.00215\pm 0.00013)~\mbox{ps}^{-1}$} from 
the measurements of $\dm_{d,s}$ and $\dg_s$. We further discuss the impact of 
$\dg_d/\dg_s$ on the CKM unitarity
triangle and present ready-to-use formulae which permit improved
predictions once updated results for the operator matrix elements
are available.
}

\title{\vskip-3cm{\baselineskip14pt
    \begin{flushleft}
      \normalsize P3H-25-107, SI-HEP-2025-27, TTP25-054 
    \end{flushleft}} \vskip1.5cm 
  Complete next-to-next-to-leading order QCD corrections to
  the decay matrix in $\boldsymbol{B}$-meson mixing  at leading power
}
\author{
  Ulrich Nierste$^a$, 
  Pascal Reeck$^a$, 
  Vladyslav Shtabovenko$^b$,
  and Matthias Steinhauser$^a$
  \\[1em]
  {\it $^a$\small\it Institut f{\"u}r Theoretische Teilchenphysik,}\\
  {\small\it Wolfgang-Gaede Stra\ss{}e 1, Karlsruhe Institute of Technology (KIT)}\\
  {\small\it 76131 Karlsruhe, Germany}  
  \\[.5em]
  {\it $^b$\small Theoretische Physik 1, Center for Particle Physics Siegen,} \\
\small\it Universit\"at Siegen, 57068 Siegen, Germany}

\date{}

\begin{document}
\maketitle
\flushbottom

%- {{{ Introduction:

\section{Introduction}
$B_d$ and $B_s$ mesons mix with their antiparticles via box diagrams
involving two $W$ bosons. As a consequence, a meson tagged as
$B_q$, where $q=d$ or $q=s$, at time $t=0$ evolves into a superposition
of $B_q$ and $\bar B_q$ for $t>0$. The time evolution of this coupled 
system reads
\begin{align}
    i \frac{\text{d}}{\text{d} t} \left( \!
\begin{array}{c}
\ds \ket{B_q (t)} \\[1mm]
\ds \ket{\,\Bbar_q (t)}
\end{array}\! \right) &  = \;  
\lt( M^q - i \frac{\Gamma^q}{2} \rt) 
\left(\!
\begin{array}{c}
\ds \ket{B_q (t)} \\[1mm]
\ds \ket{\,\Bbar_q(t)}
\end{array}\!\right) .\label{eq:tevl}
\end{align}
$M^q=M^{q\dagger}$ and $\Gamma^q=\Gamma^{q\dagger}$ are the $2\times 2$
mass and decay matrix, respectively, and \bbmq\ occurs because the
off-diagonal elements $M_{12}^q$ and $\Gamma_{12}^q$ are non-zero. The
initial conditions for the time-dependent states are
$\ket{B_q (0)} = \ket{B_q} $ and $\ket{\Bbar_q (0)} = \ket{\Bbar_q} $.
By diagonalising $M^q - i \Gamma^q/2$ one finds the mass
eigenstates $\ket{B_{q,L}(t)}$ and $\ket{B_{q,H}(t)}$ which obey
exponential decay laws,
\begin{align*}
  \ket{B_{q,L}(t)} &=\, \ket{B_{q,L}} e^{-i M_L^q t-\Gamma_L^q t/2}, \qquad &  
  \ket{B_{q,H}(t)} &=\, \ket{B_{q,H}} e^{-i M_H^q t-\Gamma_H^q t/2}.
\end{align*} 
The labels refer to ``light'' and ``heavy'' and the exponentials involve
the masses and widths of the two eigenstates.  % Thanks to $M_{11}^q=M_{22}^q$
% and $\Gamma_{11}^q=\Gamma_{22}^q$ one can perform the diagonalisation
% with the ansatz
One can perform the diagonalisation with the ansatz
\begin{align}
  \ket{B_{q,L}} &=  p \ket{B_q} + q \ket{\Bbar_q}\,, &
                                                       \ket{B_{q,H}}
  &=  p \ket{B_q} - q \ket{\Bbar_q}\,, &
\quad \mbox{where } |p|^2+|q|^2=1.
                                                         \label{eq:defpq}
\end{align}
\bbmq\ is a $\Delta B=2$ transition, meaning that the beauty quantum
number changes by two units. These transitions are characterised by
four theoretical quantities which are 
\begin{align}
  |M_{12}^q|,\qquad    |\Gamma_{12}^q|, & \qquad 
  \phi_q=\arg \lt( - \frac{\Gamma_{12}^q}{M_{12}^q}\rt), \label{eq:thp}
\end{align}
and the relative phase between $M_{12}^q$ and the amplitude of a chosen
$b$ quark decay \cite{Buras:1984pq}.  For the latter usually
$b\to c\bar c s$ is taken and within the Standard Model (SM) the
corresponding mixing-induced CP asymmetry in $B_d(t) \to J/\psi K_S$
permits the experimental determination of the angle
$\beta = \arg(-V_{cd}V_{cb}^*/(V_{td}V_{tb}^*))$ of the standard
unitarity triangle of the Cabibbo-Kobayashi-Maskawa (CKM) matrix $V$.  The
counterpart for \bbms\ is $B_s(t) \to J/\psi \phi$, measuring the angle
$\beta_s = \arg(-V_{cs}^*V_{cb}/(V_{ts}^*V_{tb})) $ of a ``squashed''
unitarity triangle in the SM. The three quantities in \eq{eq:thp}
universally affect all decays of neutral $B_q$ mesons and are related to
three phenomenological quantities as
\begin{align}
  \dm_q & \equiv \, M_H^q-M_L^q = \, 2|M_{12}^q|  {\, ,}
  \nonumber\\
   \dg_q &\equiv \, \Gamma_L^q-\Gamma_H^q
   =\,  2 |\Gamma_{12}^q| \cos \phi_q\,,
    \nonumber\\  
 a_{\rm fs}^q &\equiv \, -\frac{|\Gamma_{12}^q|}{|M_{12}^q|} \sin \phi_q
  = \mbox{Im} \frac{\Gamma_{12}^q}{M_{12}^q}\,. 
                 \label{eq:dgdmafsq}
\end{align}
Furthermore, we introduce
\begin{align}
  \frac{\dg_q}{\dm_q}
        &=\, 
          - \mbox{Re}\frac{\Gamma_{12}^q}{M_{12}^q}\,.\label{eq:dgdmafsq2}
\end{align}
The quantity $a_{\rm fs}^q$ is the CP asymmetry in flavour-specific
decays, characterised by the feature that $B_q\to f$ is allowed while
$\bar B_q\to f$ is forbidden. The corresponding CP asymmetry reads
\begin{equation}
 \frac{\gqbtf -  \gdtfb}{\gqbtf + \gdtfb} 
= a_{\rm fs}^q  + a_{\rm CP}^{\rm dir}. 
\label{afstc} 
\end{equation}
In Refs.~\cite{Nierste:2004uz,Gerlach:2025tcx} it is explained how one can
disentangle $a_{\rm fs}^q$ from the direct CP asymmetry
$ a_{\rm CP}^{\rm dir}$ by e.g.\ studying the untagged rate. In
\eqsand{eq:dgdmafsq}{afstc} we have neglected corrections of order
$|\Gamma_{12}^q/M_{12}^q|^2 \sim 10^{-5}$ or
$ \mbox{Im} [\Gamma_{12}^q/M_{12}^q] \lesssim 10^{-3}$ and will do so
throughout this paper.

$\dm_q$, which equals the \bbmq\ oscillation frequency, and the CP
phases $\beta_{(s)}$ determine the magnitude and phase of $M_{12}^q$ but are
insensitive to $\Gamma_{12}^q$. These quantities have different
sensitivities to new physics and $\Gamma_{12}^q$ is e.g.\ 
interesting to probe $b$ decays into final states
with new invisible particles \cite{Alonso-Alvarez:2021qfd} .

The theory predictions of $\dg_q$ and $a_{\rm fs}^q$ all involve the
decay matrix element $\Gamma_{12}^q$. To calculate $M_{12}^q$ or
$\Gamma_{12}^q$ one employs an operator product expansion (OPE)
separating the short-distance physics described by perturbatively
calculable Wilson coefficients from the long-distance dynamics contained
in hadronic operator matrix elements calculated by non-perturbative
methods such as lattice gauge theory. Yet 
the calculation of
$\Gamma_{12}^q$ differs from the one of  $M_{12}^q$ in two important
aspects: Firstly,  $\Gamma_{12}^q$ involves an expansion in powers of
$\lqcd/m_b$, the \emph{Heavy Quark Expansion} (HQE), where $\lqcd \sim
400\,\mev$ is the fundamental scale of QCD with new operators and
coefficients in each power of $\lqcd/m_b$~\cite{Khoze:1986fa}. Secondly, the QCD
corrections to $\Gamma_{12}^q$ are governed by $\alpha_s(m_b)$, which is
about twice as large as  $\alpha_s(m_t)$ entering the perturbative series 
for $M_{12}^q$. In this paper we present the complete NNLO corrections
to the leading-power term of $\Gamma_{12}^q$. In the phenomenological
applications we need the ratio  $\Gamma_{12}^q/M_{12}^q$ and the NLO
result for $M_{12}^q$ is sufficient to match the NNLO precision of
$\Gamma_{12}^q$.

\begin{figure}
  \centering{\includegraphics[height=3cm]{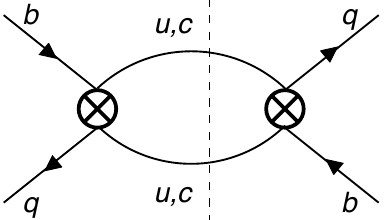}
    \hspace{2cm}
    \includegraphics[height=3cm]{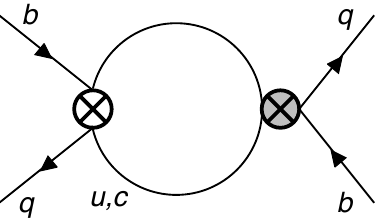}}
  
\caption{Leading-order diagrams contributing to $\Gamma_{12}^q$ with two
  current-current operators (left) and one current-current and one
  penguin operator (right). $\Gamma_{12}^q$ originates from decays into
  final states into which both $B_q$ and $\bar B_q$ can decay, indicated
  by the dashed line in the left figure.  
  \label{fig:dblo}  
 }
\end{figure}

For the calculation of $\Gamma_{12}^q$ one first employs the weak $|\Delta B|=1$
Hamiltonian ${\cal H}_{\rm eff}^{|\Delta B|=1}$ obtained by integrating
out top quark and $W$ boson,
\begin{align}
  {\cal H}_{\rm eff}^{|\Delta B|=1} 
  &=\,    {\cal H}_{\rm cc}^{|\Delta B|=1} +  {\cal H}_{\rm peng}^{|\Delta B|=1}\,,
 \label{eq:heff}
\end{align}
which describes the weak interaction in terms of effective dimension-six
operators with the two terms containing current-current and penguin
operators, respectively.  The former are four-quark operators describing
$W$-mediated tree-level decays and QCD corrections to it.
${\cal H}_{\rm peng}^{|\Delta B|=1}$ involves four additional four-quark
operators and the chromomagnetic operator.  We can safely neglect
higher-order electroweak corrections, which would add more operators to
${\cal H}_{\rm peng}^{|\Delta B|=1}$.  The Wilson coefficients of the
current-current operators have been calculated to next-to-leading order
(NLO) in Refs.~\cite{Buras:1989xd,Buras:1991jm} and to
next-to-next-to-leading order (NNLO) in Refs.~\cite{Gorbahn:2004my}. The
chromomagnetic operator is proportional to the strong coupling, so that
we will only need the corresponding NLO coefficient
\cite{Misiak:1994zw,Chetyrkin:1996vx} for our NNLO prediction of
$\Gamma_{12}^q$.
Since $\Gamma_{12}^q$ involves $|\Delta B|=2$ transitions, it is
generated at second order of  ${\cal H}_{\rm eff}^{|\Delta B|=1}$:  
\begin{equation}
  \Gamma_{12}^q = \frac{1}{2 M_{B_s}} \,\mbox{Abs}
                    \langle B_q|\, i \! \int{\rm d}^4 x \,\, T\,\,
                    {\cal H}_{\rm eff}^{\Delta B=1}(x)
                    {\cal H}_{\rm eff}^{\Delta B=1}(0)
                    |\bar{B}_q\rangle\,, 
                    \label{eq:ot}
\end{equation}
where ``Abs'' denotes the absorptive part of the matrix element.
The contribution to $\Gamma_{12}^q$ with two current-current

insertions,
\begin{equation}
  \Gamma_{12\, cc}^q = \frac{1}{2 M_{B_s}} \,\mbox{Abs}
                    \langle B_q| \, i \! \int{\rm d}^4 x \,\, T\,\,
                    {\cal H}_{\rm cc}^{\Delta B=1}(x)
                    {\cal H}_{\rm cc}^{\Delta B=1}(0)
                    |\bar{B}_q\rangle\,, 
                    \label{eq:g12c}
\end{equation}
is shown in the left diagram of \fig{fig:dblo}.
The HQE expresses $\Gamma_{12\, cc}^q$ as a linear combination of
local dimension-six $\Delta B=2$ operators whose LO Wilson coefficients are found by
matching the diagrams of \fig{fig:dblo} to the tree-level matrix
elements of these operators. For the leading power of the $1/m_b$
expansion we need two physical $\Delta B=2$ operators. Their hadronic
matrix elements are the only non-perturbative quantities need for the
predictions of $\dg_q$ and $a_{\rm fs}^q$. 

The leading-power contribution to $\Gamma_{12\, cc}^q$ has been
calculated to NLO in
Ref.~\cite{Beneke:1998sy,Beneke:2003az,Lenz:2006hd}. This calculation
involves two-loop diagrams with one extra gluon compared to
\fig{fig:dblo} as well as one-loop matrix elements of the $\Delta B=2$
operators and results in ${\cal O} (\alpha_s)$ corrections to the
desired $\Delta B=2$ matching coefficients. Subsequently NNLO corrections
have been calculated to order $\alpha_s^2 n_f$, where $n_f=5$ is the
number of active quarks,
\cite{Asatrian:2017qaz,Asatrian:2020zxa,Hovhannisyan:2022miy} and
finally completely to order $\alpha_s^2$
\cite{Gerlach:2022hoj,Gerlach:2025tcx}.  These calculations involve
expansions in the quark mass ratio $m_c/m_b$ and the result of
Ref.~\cite{Gerlach:2025tcx} with matching coefficients up to order $(m_c/m_b)^{20}$
constitutes our final NNLO prediction of $\Gamma_{12\, cc}^q$.

$ {\cal H}_{\rm peng}^{|\Delta B|=1}$ contributes less to
$\Gamma_{12}^q$ than  ${\cal H}_{\rm cc}^{|\Delta B|=1}$ because
the numerical size of the Wilson coefficients of four-quark penguin operators is smaller as compared to the four-quark current ones, and the  
chromomagnetic penguin operator is proportional to $\alpha_s$. The
second-largest contribution to $\Gamma_{12}^q$ is therefore
\begin{equation}
  \Gamma_{12\, cp}^q = \frac{1}{M_{B_s}} \,\mbox{Abs}
                    \langle B_q| \, i \! \int{\rm d}^4 x \,\, T\,\,
                    {\cal H}_{\rm cc}^{\Delta B=1}(x)
                    {\cal H}_{\rm peng}^{\Delta B=1}(0)
                    |\bar{B}_q\rangle\,, 
                    \label{eq:g12cp}
\end{equation}
subsuming the interference effects of ${\cal H}_{\rm cc}^{|\Delta B|=1}$
and ${\cal H}_{\rm peng}^{|\Delta B|=1}$, with the LO contribution
depicted in the right diagram of \fig{fig:dblo}. The NLO contribution
with a chromomagnetic operator to $\Gamma_{12\, cp}^q$ involves only a
one-loop diagram and is included in the result of
Ref.~\cite{Beneke:1998sy}.  The NLO calculation of $\Gamma_{12\, cp}^q$
including four-quark penguin operators has been performed in
Ref.~\cite{Gerlach:2021xtb} to order $(m_c/m_b)^2$. Ref.~\cite{Gerlach:2022wgb} 
contains the NNLO contribution to $\Gamma_{12\, cp}^q$ involving one
chromomagnetic operator. 

The third contribution to $\Gamma_{12}^q$ is
\begin{equation}
  \Gamma_{12\, pp}^q = \frac{1}{2 M_{B_s}} \,\mbox{Abs}
                    \langle B_q| \, i \! \int{\rm d}^4 x \,\, T\,\,
                    {\cal H}_{\rm peng}^{\Delta B=1}(x)
                    {\cal H}_{\rm peng}^{\Delta B=1}(0)
                    |\bar{B}_q\rangle\, . 
                    \label{eq:g12p}
\end{equation}
While $ \Gamma_{12\, cp}^q$ contributes to both $\dg_q$ and
$a_{\rm fs}^q$, this is not the case for $\Gamma_{12\, pp}^q$, which
does not affect $a_{\rm fs}^q$. This is because $\Gamma_{12\, pp}^q$ and
$M_{12}^q$ have the same CKM factor, so that
$\imag (\Gamma_{12\, pp}^q/M_{12}^q)=0$ in \eq{eq:dgdmafsq2}.  The
$\alpha_s n_f$ terms and the complete NLO corrections to
$ \Gamma_{12\, pp}^q$ have been calculated in
Refs.~\cite{Asatrian:2020zxa} and \cite{Gerlach:2022wgb}, respectively,
both to order $(m_c/m_b)^2$. The latter reference also contains two-loop
contributions with one or two insertions of the chromomagnetic operator,
which go beyond NNLO. Finally we mention that the $1/m_b$ power
corrections are only calculated to LO of QCD
\cite{Beneke:1996gn,Dighe:2001gc,Lenz:2006hd} and furthermore involve
so-far poorly known hadronic matrix elements \cite{Davies:2019gnp}.

In this paper, we present the complete NLO and NNLO calculation of the
leading-power terms in 
$\Gamma_{12\, cp}^q$ and  $\Gamma_{12\, pp}^q$, which includes a deep
expansion in  $m_c/m_b$ and novel three-loop results for diagrams with
two insertions of four-quark penguin operators. To appreciate the
first point we remark that $a_{\rm fs}^q \propto (m_c/m_b)^2$, so that
a precise prediction calls for an expansion of $\Gamma_{12}^q$ beyond
$ (m_c/m_b)^2 $. The NNLO prediction of $\dg_s$ in Ref.~\cite{Gerlach:2025tcx}
has a theoretical  uncertainty of the same size as the current
experimental error, so that improvement is desirable.
With the inclusion of the missing penguin contributions we
aim at decreasing this uncertainty, which we estimate from the
renormalisation scale dependence. 

\begin{figure}
\centering{\includegraphics[height=8cm]{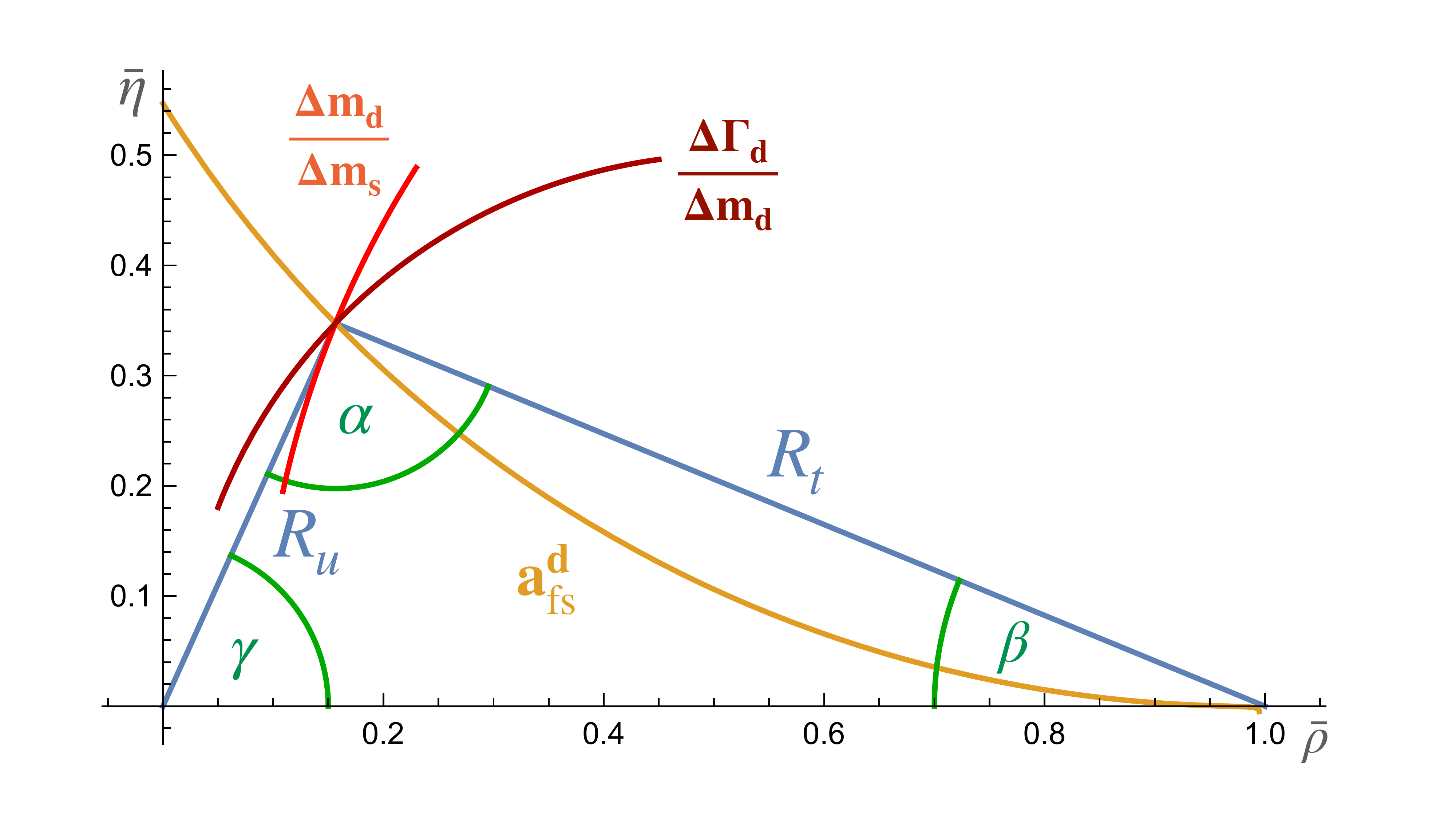}}
\caption{CKM unitarity triangle with constraints from \bbm\
  observables.\label{fig:ut}}
\end{figure}

The phenomenological role of the four quantities $\dg_q$ and $a_{\rm
  fs}^q$ with $q=d$ or $s$ is very different: The experimentally
well-determined ratio $\dg_s/\dm_s$ is essentially independent of all
CKM parameters, including the controversial $|V_{cb}| $, and therefore
directly probes the SM \cite{Gerlach:2022hoj,Gerlach:2025tcx}.
The SM prediction for $a_{\rm fs}^s$ is too small to ever be measured,
but can be two orders of magnitude larger in the presence of new physics \cite{Lenz:2006hd}.
Future precise measurements of $a_{\rm   fs}^d$ and  $\dg_d$ will
provide novel constraints of the apex $(\bar\rho,\bar\eta)$ of the CKM
unitarity triangle (UT). We
define the CKM combinations
\begin{equation}
  \lambda^q_u = V_{uq}^* V_{ub}\,,\qquad
  \lambda^q_c = V_{cq}^* V_{cb}\,,\qquad 
  \lambda^q_t = V_{tq}^* V_{tb}\,,
\end{equation}
and recall
\begin{align}
  -\frac{\lambda_u^d}{\lambda_c^d}
  &=\,  R_u e^{-i\gamma }=\, \bar\rho - i \bar\eta,
\nonumber\\%  \qquad\qquad
    -\frac{\lambda_t^d}{\lambda_c^d}
  &=\,  R_t e^{i\beta} =\, 1- \bar\rho + i \bar\eta,
  \nn
    -\frac{\lambda_u^d}{\lambda_t^d}
  &=\,  \frac{R_u}{R_t} e^{i\alpha} =\,
    1- \frac{e^{-i\beta}}{R_t}  =\, 
    %1- \frac{1-\bar\rho-i \bar\eta}{(1-\bar\rho)^2+\bar\eta^2},
    \frac{-\bar\rho (1-\bar\rho) +\bar\eta^2 + i \bar\eta}{(1-\bar\rho)^2+\bar\eta^2},
    \label{eq:utp}
\end{align}
which involves angles and sides of the UT  shown
in \fig{fig:ut}.
The constraint from $a_{\rm fs}^d$ has already been discussed in
Refs.~\cite{Beneke:2003az,Gerlach:2025tcx}, and \fig{fig:ut} further
shows the orthogonal constraint from $\dg_d/\dm_d$. The dominant
contribution to this quantity comes with the same QCD corrections as
$\dg_s/\dm_s$ and the hadronic matrix elements are the same up to
 SU(3)$_F$ breaking corrections. 
Furthermore, the CKM factor $ \real (\lambda_u^d/\lambda_t^d)$ is proportional to $\cos\alpha$ (see \eq{eq:utp}), which the global SM fit to the UT determines close to 0, so the linear and quadratic terms of this ratio in $\Delta\Gamma_d/\Delta M_d$ are negligible. Similarly, the quantity $ \real (\lambda_u^s/\lambda_t^s) $ is tiny (see, e.g., Eq.~(73) of Ref.~\cite{Gerlach:2025tcx}), so there is no substantial difference in $\Delta \Gamma/\Delta M$ across the $B_s$ and $B_d$ systems stemming from the CKM factors.
These features allow us to predict the double ratio
$\dg_d\dm_s/(\dm_d \dg_s)$ with high precision. Since three of
the four involved quantities are already measured precisely, the future
determination of $\dg_d$ will provide us with an excellent test of the
SM from the double ratio.

The paper is organised as follows: In the next section we present the
effective Hamiltonian and the operators used in this paper. Furthermore, we
discuss in detail the numerical result for the $1/m_b$-suppressed contributions as extracted from the literature.
Section~\ref{sec::phen} is dedicated to phenomenology
and discusses the quantities $\Delta\Gamma_q$, $a_\text{fs}^q$ and $\Delta\Gamma_q/\Delta M_q$
for the $B_s$ and $B_d$ systems. We provide NNLO corrections
including a detailed uncertainty analysis. Furthermore, NNLO predictions
for the double ratio $(\Delta\Gamma_s/\Delta M_s) / (\Delta\Gamma_d/\Delta M_d)$
are discussed. Finally, we construct $\Delta\Gamma_d / \Delta\Gamma_s$
and discuss its complementary to $a_{\rm fs}^d$ for providing
constraints on the CKM unitarity triangle.
We summarise our findings in Section~\ref{sec::concl}.

\section{General formalism}
The $|\Delta B|=1$ Hamiltonian in the CMM operator
basis \cite{Chetyrkin:1997gb} reads
\begin{eqnarray}
  \mathcal{H}_{\textrm{eff}}^{|\Delta B|=1} 
  &=&   \frac{4G_F}{\sqrt{2}}  \left[
      -\, \lambda^q_t \Big( \sum_{i=1}^6 C_i Q_i + C_8 Q_8 \Big) 
      - \lambda^q_u \sum_{i=1}^2 C_i (Q_i - Q_i^u) \right. \nn
  && \phantom{\frac{4G_F}{\sqrt{2}} \Big[}
      \left.
      +\, V_{uq}^\ast V_{cb} \, \sum_{i=1}^2 C_i Q_i^{cu} 
      + V_{cq}^\ast V_{ub} \, \sum_{i=1}^2 C_i Q_i^{uc} 
      \right]
      + \mbox{h.c.}\,,
      \label{eq::HamDB1}
\end{eqnarray}
where $G_F$ is the Fermi constant and 
\begin{align}
  Q^u_1 &=\,  \bar{q}_L \gamma_{\mu} T^a u_L\;\bar{u}_L     \gamma^{\mu} T^a b_L\,,&
  Q^u_2 &=\,  \bar{q}_L \gamma_{\mu}     u_L\;\bar{u}_L     \gamma^{\mu}     b_L\,,\nonumber\\
  Q^{cu}_1 &=\,  \bar{q}_L \gamma_{\mu} T^a u_L\;\bar{c}_L     \gamma^{\mu} T^a b_L\,,&
  Q^{cu}_2 &=\,  \bar{q}_L \gamma_{\mu}     u_L\;\bar{c}_L     \gamma^{\mu}     b_L\,,\nonumber\\
  Q^{uc}_1 &=\,  \bar{q}_L \gamma_{\mu} T^a c_L\;\bar{u}_L     \gamma^{\mu} T^a b_L\,,&
  Q^{uc}_2 &=\,  \bar{q}_L \gamma_{\mu}     c_L\;\bar{u}_L
             \gamma^{\mu}     b_L\,, \nonumber\\
  Q_1   &=\,  \bar{q}_L \gamma_{\mu} T^a c_L\;\bar{c}_L     \gamma^{\mu} T^a b_L\,,& 
  Q_2   &=\,  \bar{q}_L \gamma_{\mu}     c_L\;\bar{c}_L     \gamma^{\mu}     b_L\,,\nonumber\\
  Q_3   &=\,  \bar{q}_L \gamma_{\mu}     b_L \sum_{q^\prime} \bar{q}{}^\prime\gamma^{\mu}     q^\prime\,,& 
  Q_4   &=\,  \bar{q}_L \gamma_{\mu} T^a b_L \sum_{q^\prime} \bar{q}{}^\prime\gamma^{\mu} T^a q^\prime\,,\nonumber\\
  Q_5   &=\,  \bar{q}_L \gamma_{\mu_1}
            \gamma_{\mu_2}
            \gamma_{\mu_3}    b_L\sum_{q^\prime} \bar{q}{}^\prime \gamma^{\mu_1} 
            \gamma^{\mu_2}
            \gamma^{\mu_3}     q^\prime\,,& 
  Q_6   &=\,  \bar{q}_L \gamma_{\mu_1}
            \gamma_{\mu_2}
            \gamma_{\mu_3} T^a b_L\sum_{q^\prime} \bar{q}{}^\prime \gamma^{\mu_1} 
            \gamma^{\mu_2}
            \gamma^{\mu_3} T^a q^\prime\,,\nonumber\\
  Q_8  &=\,   \frac{g_s}{16\pi^2} m_b \, \bar{q}_L \sigma^{\mu \nu} T^a
           b_R \, G_{\mu\nu}^a\, .
           \label{operators}
\end{align}
Here $q_L=P_Lq$ with $P_L=(1-\gamma_5)/2$,
$\sigma^{\mu \nu} = i[\gamma^\mu,\gamma^\nu]/2$, and $T^a$ is the
generator of SU(3)$_c$.

It helps to write $\Gamma_{12}^q$ as
\begin{align}
  \Gamma_{12}^q
   &= - (\lambda_c^q)^2\Gamma^{cc}_{12} 
        - 2\lambda_c^q\lambda_u^q \Gamma_{12}^{uc} 
        - (\lambda_u^q)^2\Gamma^{uu}_{12} \label{eq:gam12a} \\
    &= -(\lambda_t^q)^2 \left[
      \Gamma_{12}^{cc} 
      + 2 \frac{\lambda_u^q}{\lambda_t^q}\left(\Gamma_{12}^{cc}-\Gamma_{12}^{uc}\right)
      + \left(\frac{\lambda_u^q}{\lambda_t^q}\right)^2 
      \left(\Gamma_{12}^{uu}+\Gamma_{12}^{cc}-2\Gamma_{12}^{uc}\right)
      \right]
      \, . \numberthis
        \label{eq::Gam12}
\end{align}
with real coefficients $\Gamma_{12}^{ab}$ multiplying the three possible
CKM factors. The current-current contributions in \fig{fig:dblo} match
\eq{eq:gam12a} such that $a,b=u,c$ correspond to the quarks on the
internal lines.  In the step from \eq{eq:gam12a} to \eq{eq::Gam12} we
have used $\lambda_c^q=-\lambda_t^q-\lambda_u^q$ to get an expression best suited to normalise $ \Gamma_{12}^q$ to $ M_{12}^q$, which
is proportional to $(\lambda_t^q)^2$.  The penguin operators in
${\cal H}_{\rm eff}^{|\Delta B|=1}$ are proportional to $\lambda_t^q$.
Terms quadratic in the penguin coefficients 
contribute to all three $\Gamma^{ab}$'s in \eq{eq:gam12a}
but only to the first term in the square bracket of \eq{eq::Gam12}. 
The second term in this square bracket receives contributions from 
the interference of penguin and current-current operators.

The individual contributions in \eq{eq::Gam12} evaluate to \cite{Gerlach:2025tcx}
\begin{equation}
        \Gamma_{12}^{ab} 
        = \frac{G_F^2 m_b^2}{24\pi M_{B_q}} \left[ 
        H^{ab}(\bar z)   \langle B_q|Q|\bar{B}_q \rangle
        + \widetilde{H}^{ab}_S(\bar z)  \langle B_q|\widetilde{Q}_S|\bar{B}_q \rangle
        \right]
        +    %\mathcal{O}(\Lambda_{\rm QCD}/m_b) \,,
         \Gamma_{12,\, 1/m_b}^{ab} 
        \label{eq::Gam^ab}
\end{equation}
with two  matrix elements of dimension-six $\Delta B=2$ operators,  
\begin{align}
        Q &= 4\,(\bar{s}^c \gamma^\mu P_L b^c) \; (\bar{s}^d \gamma_\mu
        P_L b^d)\,, \nonumber\\
        \widetilde{Q}_S &= 4\,(\bar{s}^c P_R b^d)\; (\bar{s}^d P_R
        b^c)\,, % \\
    %Q_S &= 4\,(\bar{s}^c P_R b^c) \;(\bar{s}^d P_R
     %   b^d)\,,\numberthis 
 \label{eq::opDB2}
\end{align}
where $P_R=(1+\gamma_5)/2$ and $c,d$ are colour indices. The mass ratio $\bar z=[m_c({{\mu_b}})/m_b({\mu_b})]^2$ 
involves the charm and bottom masses in the $\ov{\rm MS}$ scheme, both defined in five-flavour QCD, $M_{B_q}$ is the $B_q$ meson mass, and $G_F$ is the Fermi constant. The calculation of the Wilson coefficients 
$H^{ab}(\bar z)  $ and $\widetilde{H}^{ab}_S(\bar z)$ for $ab=cc,uc,uu$ at NNLO of QCD is the subject of this paper. 

$\Gamma_{12,\, {1/m_b}}^{ab}$ comprises power corrections of order $\lqcd/m_b$ which have been calculated at 
LO of QCD in Ref.~\cite{Beneke:1996gn}, NLO results are not available yet. As a first step  to NLO, 
the mixing of dimension-seven operators into dimension-six operators under renormalization has been recently calculated in Ref.~\cite{
Hovhannisyan:2025zev}.  Since LO $\lqcd/m_b$-terms will finally dominate the theoretical uncertainties of our predictions, we present $\Gamma_{12,\, {1/m_b}}^{ab}$ here in a way which allows the reader 
to include future improvements of the calculation of the hadronic matrix elements entering $\Gamma_{12,\, {1/m_b}}^{ab}$. $\Gamma_{12,\, {1/m_b}}^{ab}$ involves seven dimension-seven operators commonly denoted by 
$R_0$, $R_k$, and $\widetilde R_k$ with $k=1,2,3$. Their definitions can be found in Refs.~\cite{Beneke:1996gn,Lenz:2006hd,Davies:2019gnp}. 

Defining $r_j^q \equiv \bra{B_q} R_j \ket{\bar B_q}$ and
$\widetilde r_j^q \equiv \bra{B_q} \widetilde R_j \ket{\bar B_q}$
one finds
\begin{align}
        \Gamma_{12,\, 1/m_b}^{ab} 
        &=\, \frac{G_F^2 m_b^2}{24\pi M_{B_q}} \lt[ 
         g_0^{ab} r_0^q \, + \,
   \sum_{j=1}^3 \lt[ g_j^{ab} r_j^q + 
       \widetilde{g}_j^{ab} \widetilde r_j^q
       \rt]
      \rt] \label{ga12m}   
\end{align}
with the coefficients $g_j^{ab}$, $\widetilde g_j^{ab}$ (subsuming the
results of Refs.~\cite{Beneke:1996gn,Dighe:2001gc,Beneke:2003az}) defined
in Ref.~\cite{Lenz:2006hd}.

Estimating the uncertainties of $g_j^{ab}$, $\widetilde g_j^{ab}$ from 
their $\mu$-dependence in the interval $[\SI{2.1}{GeV},\SI{8.4}{GeV}]$, we quote our central values for the scale $\mu_1=\SI{4.2}{GeV}$ and the matching scale $\mu_0=\SI{165}{GeV}$ of the weak effective Hamiltonian, finding
\begin{align}
        \Gamma_{12,\, 1/m_b}^{cc} 
        &=\, \frac{G_F^2 m_b^2}{24\pi M_{B_q}} \big[ 
          %(-0.37 \pm 0.08) r_0^q +
          {(-0.42 \pm 0.07) r_0^q +}
          %(0.75 \pm 0.17) r_1^q - (2.45 \pm 0.23 ) \widetilde r_1^q
          {(0.84 \pm 0.14) r_1^q - (2.58 \pm 0.27 ) \widetilde r_1^q}
           \nn
          %&\quad + ( 0.76 \pm 0.17)  r_2^q - (2.49 \pm 0.23) \widetilde r_2^q + 
          &\quad + {( 0.85 \pm 0.15)  r_2^q - (2.62 \pm 0.28) \widetilde r_2^q + }
            %(0.03\pm 0.01) r_3^q - (0.08 \pm 0.01) \widetilde r_3^q \big] \label{eq:g12cc1}\\[2mm]
            {(0.03\pm 0.01) r_3^q - (0.09 \pm 0.01) \widetilde r_3^q} \big] \label{eq:g12cc1}\\[2mm]
  &=\lt\{ \begin{array}{ll} \ds
            - \lt( \frac{m_b}{4.48\gev}\rt) ^2 \,   
            % 5.086 \cdot 10^{-12} \gev
            %7.73\,\mbox{ps}^{-1}
            {7.77}\,\mbox{ps}^{-1}
            %\, (1\pm 0.13)_{\rm scale} \,
            \, {(1\pm 0.09)_{\rm scale} }\,
            (1\pm 0.49)_{\rm mat.\ el.}
            &\quad\mbox{for }q=s \\[2mm] \ds
            - \lt( \frac{m_b}{4.48\gev}\rt) ^2 \,
            %4.95 \,\mbox{ps}^{-1} \,
            {5.18} \,\mbox{ps}^{-1} \,
             %(1\pm 0.13)_{\rm scale} \,
             {(1\pm 0.10)_{\rm scale} }\,
            (1\pm 0.52)_{\rm mat.\ el.}
            &\quad\mbox{for }q=d 
          \end{array}   \rt. \label{eq:g12cc2}
\end{align}
\begin{align}
  2\big(
    \Gamma_{12,\, 1/m_b}^{cc} - & \Gamma_{12,\, 1/m_b}^{uc}\big) = \nn
  &\quad \frac{G_F^2 m_b^2}{24\pi M_{B_q}} \big[ 
    %( 0.007 \pm 0.002) r_0^q - 
    {( 0.007 \pm 0.001) r_0^q - }
    %( 0.013\pm 0.003) r_1^q +  ( 0.043\pm  0.004) \widetilde r_1^q
    {( 0.015\pm 0.003) r_1^q +  ( 0.046\pm  0.005) \widetilde r_1^q}
    \nn
  &\quad \phantom{\frac{G_F^2 m_b^2}{24\pi M_{B_q}}} + ( 0.001 \pm 0.000)  r_2^q - (0.004 \pm 0.000) \widetilde r_2^q + 
    %( 0.029 \pm 0.007) r_3^q \nn
    {( 0.032 \pm 0.005) r_3^q} \nn
  &\quad \phantom{\frac{G_F^2 m_b^2}{24\pi M_{B_q}}}
    %- ( 0.094 \pm 0.009) \widetilde r_3^q \big] \nn
    {- ( 0.099 \pm 0.010) \widetilde r_3^q} \big] \nn
  &  =\,
 - \lt( \frac{m_b}{4.48\gev}\rt) ^2 \,   
    %0.2135 \,\mbox{ps}^{-1}
    {0.2099} \,\mbox{ps}^{-1}
    %(1\pm 0.19)_{\rm scale} \,
    {(1\pm 0.07)_{\rm scale}} \,
    (1\pm 0.33)_{\rm mat.\ el.}\quad\mbox{for }q=d
    \label{eq:g12ccuc2}
\end{align}
\begin{align}
  \Gamma_{12,\, 1/m_b}^{uu} + &
    \Gamma_{12,\, 1/m_b}^{cc} -  2 \Gamma_{12,\, 1/m_b}^{uc} = \nn
  &\quad \frac{G_F^2 m_b^2}{24\pi M_{B_q}} \big[
     ( 0.001 \pm 0.000) r_0^q - 
    ( 0.001\pm 0.000) r_1^q +  ( 0.004\pm  0.000) \widetilde r_1^q
    \nn
  &\quad \phantom{\frac{G_F^2 m_b^2}{24\pi M_{B_q}}} +
    ( 0.001 \pm 0.000)  r_2^q - (0.002 \pm 0.000) \widetilde r_2^q +
    ( 0.004 \pm 0.001) r_3^q \nn
  &\quad \phantom{\frac{G_F^2 m_b^2}{24\pi M_{B_q}}}
    %- ( 0.012 \pm 0.001) \widetilde r_3^q \big] \nn
    {- ( 0.013 \pm 0.001) \widetilde r_3^q} \big] \nn
    &  =\,
 - \lt( \frac{m_b}{4.48\gev}\rt) ^2 \,   
    %0.0030 \,\mbox{ps}^{-1}
    {0.0031} \,\mbox{ps}^{-1}
    %(1\pm 0.16)_{\rm scale} \,
    {(1\pm 0.08)_{\rm scale}} \,
    (1\pm 0.33)_{\rm mat.\ el.}\quad\mbox{for }q=d
    \label{eq:g12ccucuu2}
\end{align}    
The numerical result in \eq{eq:g12cc2} is found with the hadronic matrix
elements calculated for $q=s$ in Ref.~\cite{Davies:2019gnp}; their
uncertainties combine to the value indicated by ``mat.~el.'' which
dominates over the scale uncertainty from the coefficients in
\eq{eq:g12cc1}. The small scale uncertainties in the last lines of Eqs.~(\ref{eq:g12cc2}) to~(\ref{eq:g12ccucuu2}) result from partial cancellations between the different terms
and underestimate the uncertainty from missing higher-order terms of the perturbation series. 
Still, even if one adds the quoted scale uncertainties of the coefficients in quadrature, 
one finds the uncertainty stemming from matrix elements significantly larger.

There are only five independent matrix elements due to
relations like $r_2^q=-\widetilde r_2^q (1+ {\cal O}(1/m_b,\alpha_s))$.
Those matrix elements which are not available for $q=d$ are obtained by
rescaling their $q=s$ counterparts with
$f_{B_d}^2 M_{B_d}^2/(f_{B_s}^2 M_{B_s}^2)$ and in the case of $R_1$
and $\widetilde R_1$ also by $m_d/m_s$. The largest corrections to $\Delta \Gamma_q$ 
stem from $\Gamma_{12,\, 1/m_b}^{cc}$ in \eq{eq:g12cc2}, which is
dominated by $r_2^q$, $\widetilde r_2^q$, and $r_0^q$. The terms
in Eqs.~\eqref{eq:g12ccuc2} and \eqref{eq:g12ccucuu2} give negligible contributions to $\dg_q$. For
$a_{\rm fs}^q$ one instead needs the second term of~\eq{eq::Gam12},
because the first term does not contribute to
$\imag (\Gamma_{12}^q/M_{12}^q)$. Since better lattice results will be
available by the time experiments can probe the SM prediction for
$a_{\rm fs}^s$, we only provide numerical results for $q=d$ in
\eqsand{eq:g12ccuc2}{eq:g12ccucuu2}, both of which are dominated by
$\widetilde r_3^d$ and $r_3^d= \widetilde r_3^d +\widetilde r_2^d/2$.

\section{\label{sec::tech}Amplitude generation}

In this section, we highlight some of the technical details of the amplitude generation. Since $\Gamma_{12}^q$ is computed using the matching coefficients of the $\Delta B = 2$ transition operator, the mixing amplitude is calculated from the $\Delta B = 1$ Hamiltonian as well as the $\Delta B = 2$ transition operator separately. In both cases, we use \texttt{QGRAF}~\cite{Nogueira:1991ex} to generate the amplitudes, \texttt{tapir}~\cite{Gerlach:2022qnc} to generate the topology files and insert Feynman rules, and \texttt{exp}~\cite{Harlander:1998cmq, Seidensticker:1999bb} to map the amplitudes to the corresponding topology. The
amplitudes  are then evaluated using \texttt{calc}, an in-house program written in \texttt{FORM}~\cite{Vermaseren:2000nd, Tentyukov:2007mu, Kuipers:2012rf, Ruijl:2017dtg} which makes use of  \texttt{color}~\cite{vanRitbergen:1998pn} to efficiently evaluate the QCD colour factors. The spinor structures are treated using a projector routine~\cite{Reeck:2024iwk}, which is integrated into \texttt{calc} and resolves tensor products of two spin lines with up to eleven $\gamma$ matrices each. In a separate calculation, we use tensor reduction as a cross-check for this step~\cite{Reeck:2024iwk}. The resulting scalar integrals are reduced to masters using \texttt{Kira}~\cite{Maierhofer:2017gsa,Klappert:2020nbg,Lange:2025fba}, which are then evaluated using the ``expand and match'' approach~\cite{Fael:2021kyg, Fael:2022rgm, Fael:2022miw, Fael:2023zqr}. For more details on the usage of the individual parts of the toolchain, see Ref.~\cite{Reeck:2024iwk}.

\begin{figure}[t]
    \centering
    \includegraphics[scale=0.55]{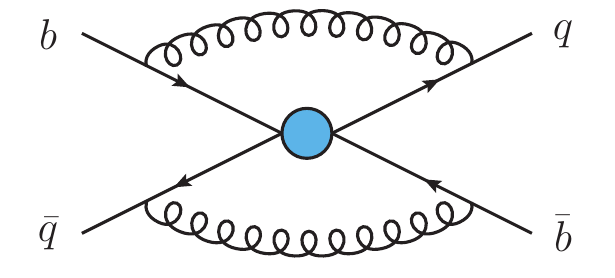} \quad
    \includegraphics[scale=0.55]{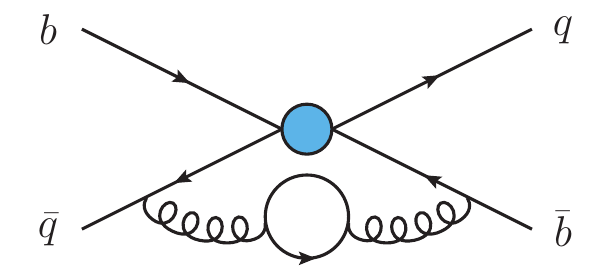}
    \caption{\label{fig::DB2}Two-loop Feynman diagrams of the $\Delta B = 2$ transition operator. The blue dot stands for the insertion of one of the
      operators $Q$, $\widetilde{Q}_S$ or $R_0$.
      }
\end{figure}

For the $\Delta B = 2$ transition operator, two-loop diagrams are calculated to obtain the NNLO amplitude, see Fig.~\ref{fig::DB2}. The % amplitude of the transition operator 
$\Delta B=2$ side of the matching equation
is the same (except for additional evanescent operators) as in Ref.~\cite{Gerlach:2025tcx} since it does not depend on the type of $\Delta B = 1$ operators used on the other side of the matching. For a more detailed discussion including the $1/m_b$-suppressed operator $R_0$, see Ref.~\cite{Gerlach:2025tcx}.

\begin{figure}[t]
    \centering
    \raisebox{0.35cm}{\includegraphics[scale=0.55]{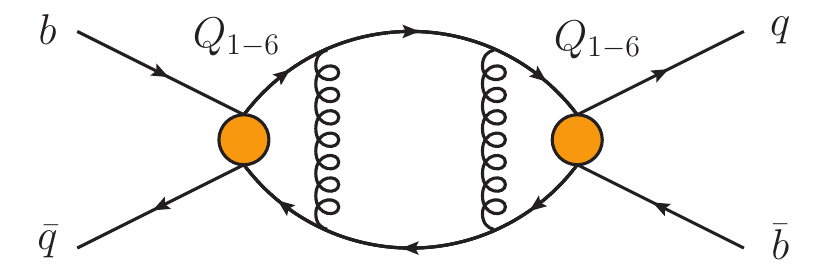}} \quad
    \includegraphics[scale=0.55]{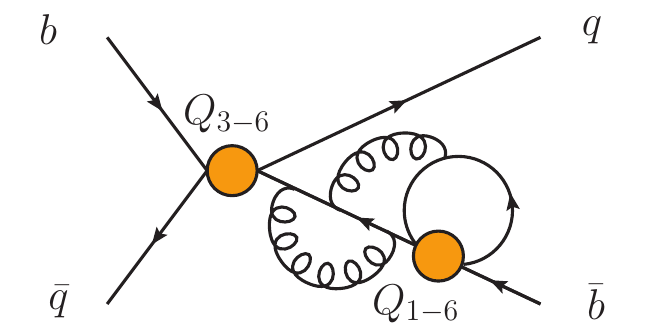}
    \caption{\label{fig::DB1}Three-loop Feynman diagrams with two $\Delta B = 1$ insertions shown in orange.
      }
\end{figure}

When calculating the mixing amplitude from two effective $\Delta B = 1$ operators, three-loop diagrams like the ones shown in Fig.~\ref{fig::DB1} are computed. The diagram type shown on the right is special because it is one-particle reducible with a flavour-changing self-energy (FCSE) on one of the external legs. While the QCD self-energies on external legs can be omitted
in the calculation, since the QCD wave function renormalisation cancels from 
the Wilson coefficients, 
the corresponding diagrams with an FCSE must be included in the amplitude. These diagrams first appear at NLO when considering insertions of one or more penguin operators.

The calculation of the three-loop amplitudes with up to two penguin operators is challenging from the perspective of the appearing spinor structures. Up to eleven $\gamma$ matrices appear on each of the two spin lines, requiring the most complicated projectors constructed in Ref.~\cite{Reeck:2024iwk}. For the case of one current-current operator and one penguin operator, we have carried out an independent cross-check using tensor integrals and Feynman gauge in the limit of $m_c=0$. 
This result depends on nine new three-loop on-shell master integrals (having at most six lines) that where not contained in the set of 23 masters calculated in Ref.~\cite{Gerlach:2022hoj}. On the level of the unrenormalised three-loop amplitude we found full agreement with the calculation using projectors and an expansion in $z=m_c^2/m_b^2$.

The matching coefficients computed in this work are available in computer-readable format up to $z^{10}$ on the website~\cite{progdata}

\section{\label{sec::phen}Phenomenology}

\begin{table}[t]
  \begin{center}
  {
    \renewcommand{\arraystretch}{1.2}
{\scalefont{0.8}
    \begin{tabular}{rclc | rclc}
      \hline 
      $\alpha_s(M_Z)$ &=& $0.1180 \pm 0.0009$ & \cite{ParticleDataGroup:2024cfk}
      &
      $\overline{m}_c(3~\mbox{GeV})$ &=& $(0.993\pm 0.008)$~GeV & \cite{Chetyrkin:2017lif}
      \\
      $m_t^{\rm pole}$ &=& $(172.4\pm 0.7)$~\mbox{GeV} & \cite{ParticleDataGroup:2024cfk} 
      &                                                 
      $\overline{m}_b(\overline{m}_b)$ &=& $(4.163\pm 0.016)$~GeV & \cite{Chetyrkin:2010ic} 
      \\
      $M_{B_s}$ &=& $(5366.88 {\pm 0.14})$~\mbox{MeV} & \cite{ParticleDataGroup:2024cfk} 
      &
      $M_{B_d}$ &=& $(5279.64 {\pm 0.12})$~\mbox{MeV} & \cite{ParticleDataGroup:2024cfk} 
      \\
      $B_{B_s}$ &=& $0.813\pm0.034$ & \cite{Dowdall:2019bea} 
      &
      $B_{B_d}$ &=& $0.806\pm0.041$ & \cite{Dowdall:2019bea}  
      \\
      $\widetilde{B}^\prime_{S,B_s}$ &=& $1.31\pm0.09$ & \cite{Dowdall:2019bea} 
      &
      $\widetilde{B}^\prime_{S,B_d}$ &=& $1.20\pm0.09$ & \cite{Dowdall:2019bea} 
      \\
      $f_{B_s}$ &=& $(0.2303\pm0.0013)$~GeV & \cite{Bazavov:2017lyh,Hughes:2017spc,ETM:2016nbo,Dowdall:2013tga} 
      &
      $f_{B_d}$ &=& $(0.1905\pm0.0013)$~\mbox{GeV} & \cite{Bazavov:2017lyh,Hughes:2017spc,ETM:2016nbo,Dowdall:2013tga} 
      \\
      $\xi$ &=& $\SI{1.216(16)}{}$ & \cite{Dowdall:2019bea} &
      $\frac{f_{B_s}}{f_{B_d}} $ &=& $\SI{1.2109(39)}{}$ & \cite{Bazavov:2017lyh}\\
      \hline 
    \end{tabular}
}
    }
  \end{center}
  \caption{\label{tab::input}Input parameters for the numerical analysis.  The
    quoted $m_t^{\rm pole}$ corresponds to {
      $m_t(m_t)=(162.6 \pm 0.7)\,\gev$} in the $\overline{\rm MS}$ scheme. We
    use the values for $B_{B_q}=B_{B_q}(\mu_2)$ and $\widetilde{B}^\prime_{S,B_q}=\widetilde{B}^\prime_{S,B_q}(\mu_2)$ with
    $\mu_2=m_b^{\rm pole}$. For our NNLO results in the pole scheme we convert the quoted $m_b(m_b)$ 
    from the $\ov{\rm MS}$ scheme with the two-loop formula to $m_b^{\rm pole}=\SI{4.758}{GeV}$. For the 
    PS scheme predictions we use $m_b^{\rm PS}=4.480\gev$ corresponding to a factorisation scale 
    of $\mu_f=2\gev$. The quoted central values for $m_{c,b}$ imply $\bar z(\SI{4.2}{GeV})=(\overline{m}_c(\SI{4.2}{GeV})/\overline{m}_b(\SI{4.2}{GeV}))^2={0.049540}$, where the charm mass is calculated for five active quark flavours. The quantity $\xi$ is defined in Eq.~(\ref{eq:xi_def}).}
%    ~\\[-5mm]
%\hrule
\end{table}

In this section, we present our result for the $B$ mixing observables as well as their implications for the determination of the CKM unitarity triangle and new physics searches. 
The input parameters used in our phenomenological analysis are given in Tab.~\ref{tab::input}, see also Section~5 of Ref.~\cite{Gerlach:2025tcx}.
In all results presented here, the scale variation is carried out between $\SI{2.1}{GeV}$ and $\SI{8.4}{GeV}$ for the purpose of determining the scale uncertainty. This is quoted for the leading and sub-leading $1/m_b$ terms separately in most cases. The uncertainties from the other input parameters are clustered in three groups: the bag parameters $B, \widetilde{B}_S$; the matrix elements of the $1/m_b$ suppressed operators, and all remaining input variables.

\subsection{$\Delta\Gamma_q$ and $\Delta\Gamma_q/\Delta M_q$}

We start with the quantities
$\Delta\Gamma$ and $\Delta\Gamma/\Delta M$ and discuss them separately for the $B_s$ and $B_d$ system. In all cases, we present results in three renormalisation schemes for the overall factor $m_b^2$.
For our final numerical prediction we choose a combination of the $\overline{\rm MS}$ and PS scheme. Results for the pole scheme are only provided for comparison.

\subsubsection{$B_s$ system}

In the three renormalisation schemes our predictions
for $\Delta\Gamma_s/\Delta M_s$
are given by
\begin{align}
  \frac{\Delta \Gamma_s}{\Delta M_s}
  ~=&~ \left(
      {{3.98^{+0.49}_{-0.53}}_{\textrm{scale}}}
      {{{}^{+0.09}_{-0.19}}_{\textrm{scale, $1/m_b$}} } 
%      {\ms\pm0.14}_{\textrm{scale, $1/m_b$}}  
      {\pm 0.11_{B\widetilde{B}_S}} \rt. 
%\nonumber\\
%  &~
    \lt. \; {\pm 0.78_{1/m_b}} { \pm 0.06_{\textrm{input}}}\right) \times
    10^{-3}\ (\textrm{pole})\,, \nonumber\\
      \frac{\Delta \Gamma_s}{\Delta M_s} 
  ~=&~ \left(
      {{4.45^{+0.19}_{ -0.40}}_{\textrm{scale}}}
      {{{}^{+0.09}_{-0.19}}_{\textrm{scale, $1/m_b$}} } 
%    {\pm0.14}_{\textrm{scale, $1/m_b$}} 
      {\pm  0.12_{B\widetilde{B}_S}} \rt. 
%\nonumber\\
%  &~
  \lt. \;     {\pm 0.78_{1/m_b} } {\pm 0.05_{\textrm{input}}}\right) \times 10^{-3}\ (\overline{\textrm{MS}})\,,
   \nonumber\\
  \frac{\Delta \Gamma_s}{\Delta M_s} ~=&~ 
 \left( {{ 4.38^{+0.33}_{ -0.35}}_{\textrm{scale}} }
      {{{}^{+0.09}_{-0.19}}_{\textrm{scale, $1/m_b$}} } 
%    {\pm0.14}_{\textrm{scale, $1/m_b$}} 
      {\pm 0.12_{B\widetilde{B}_S} } \rt. 
%\nonumber\\
%  &~
  \lt. \;{\pm 0.78_{1/m_b}}
          {\pm 0.05_{\textrm{input}}}\right) \times 10^{-3}\ (\textrm{PS})\,,
          \label{eq::dGdM}
\end{align}
where the central value is obtained at the scale $\mu_1=\mu_c=\mu_b=\SI{4.2}{GeV}$.

\begin{figure}[t]
  \begin{center}
      \includegraphics[width=0.8\textwidth]{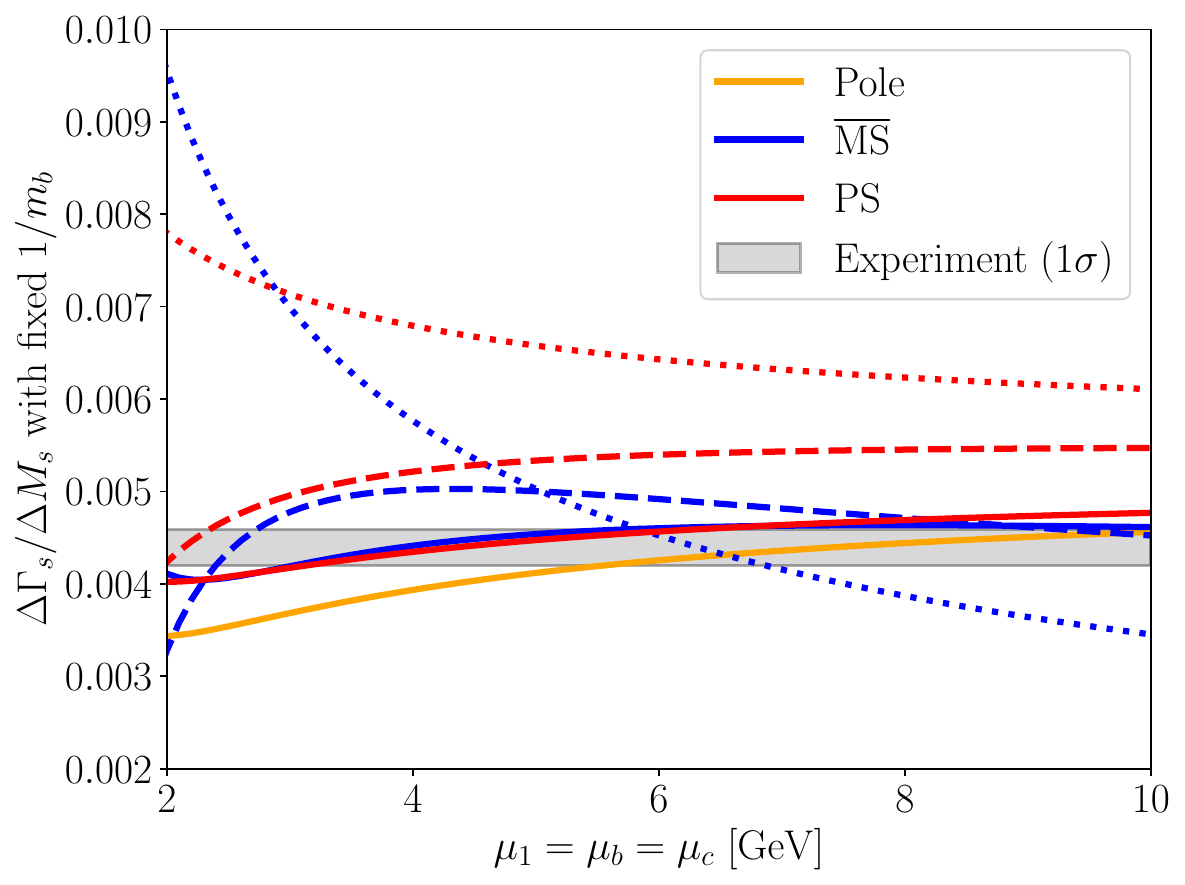}
  \end{center}
  \caption{\label{fig::DelGam_mu1} 
      Renormalisation scale dependence at LO (short dashes),
      NLO (long dashes) and NNLO (solid) for $\Delta\Gamma_s/\Delta M_s$. The plot shows the scale variation of the leading-power terms where the scales $\mu_1=\mu_b=\mu_c$ are varied together. 
      The grey band shows the experimental value of \eq{eq:dgsexp}.
      }
%~\\[-5mm]
%\hrule
\end{figure}

\begin{figure}[t]
  \begin{center}
      \includegraphics[width=0.8\textwidth]{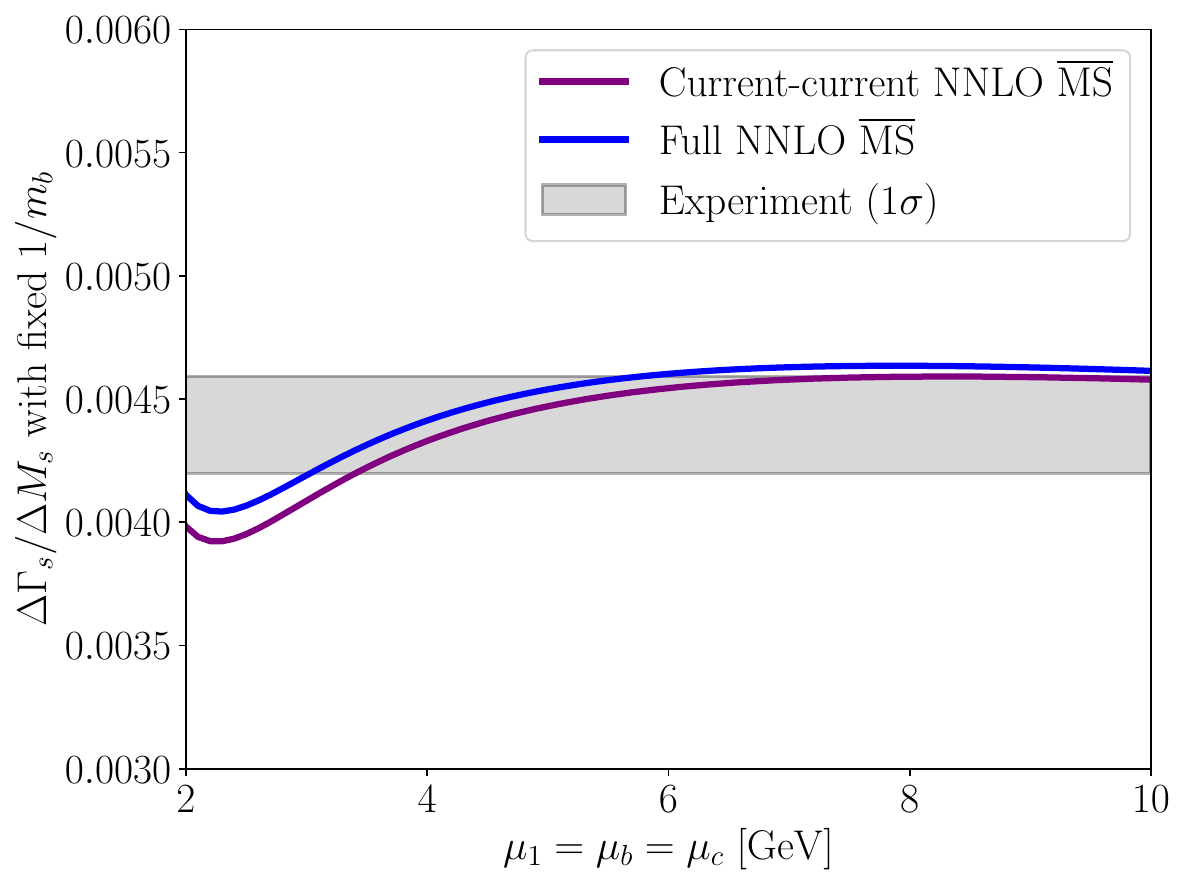}
  \end{center}
  \caption{\label{fig:dGdG_comparison} 
      Renormalisation scale dependence in the $\overline{\textrm{MS}}$ scheme for $\Delta\Gamma_s/\Delta M_s$ with and without penguin operators at NNLO. Note that the current-current result contains a shallow expansion up to $z^1$ for the NLO penguin contributions whereas the full NNLO result encompasses a deeper expansion up to $z^{10}$ at NLO too. The scale variation shown is of the leading-power terms where the scales $\mu_1=\mu_b=\mu_c$ are varied together.
      }
%~\\[-5mm]
%\hrule
\end{figure}

It is worth noting that the inclusion of the penguin operators up to NNLO results in a reduction of the symmetrised scale uncertainty by about $8\%$ on average compared to Ref.~\cite{Gerlach:2025tcx}. 
This is visualised in Fig.~\ref{fig:dGdG_comparison}.

The results presented here can be directly compared with Ref.~\cite{Gerlach:2025tcx} to gauge the impact of a deeper expansion in $z=m_c^2/m_b^2$ at NLO for the penguin contributions as well as the inclusion of the penguin operators at NNLO. 
For the $\overline{\rm MS}$ scheme, we note that for the NLO penguin contributions, i.e.~those proportional to $C_{1-6}\times C_{3-6}$, the inclusion of all charm mass terms amounts to $30\%$ of the sum of the $z^0$ and $z^1$ terms.
At NNLO, the penguin operators contribute about $12\%$ to the total NNLO part. As a result, the central value including the $1/m_b$ corrections in the $\overline{\text{MS}}$ and PS schemes is changed by about $2\%$.

Combining the experimental
result~\cite{LHCb:2021moh}
\begin{equation}
  \Delta M_s^{\rm exp} = 17.7656 \pm 0.0057~\mbox{ps}^{-1}\,
  \label{eq:dMs_exp}
\end{equation}
with the results in Eq.~(\ref{eq::dGdM}) we obtain precise predictions for $\Delta\Gamma_s$: 
\begin{align}
  \Delta \Gamma_s
  ~=&~ \left(
      {{7.06^{+0.88}_{-0.94}}_{\textrm{scale}}}
      {{}^{+0.16}_{-0.34}}_{\textrm{scale, $1/m_b$}}
    %{\pm0.25}_{\textrm{scale, $1/m_b$}} 
      \pm 0.19_{B\widetilde{B}_S}
    \; \pm 1.39_{1/m_b} \pm 0.10_{\textrm{input}}
    \right) \times
    10^{-2}~\mbox{ps}^{-1}\ 
      (\textrm{pole})\,, \nonumber\\
      \Delta \Gamma_s
  ~=&~ \left(
      {{7.90^{+0.34}_{-0.71}}_{\textrm{scale}}}{{}^{+0.16}_{-0.34}}_{\textrm{scale, $1/m_b$}}
    %{\pm0.25}_{\textrm{scale, $1/m_b$}} 
      \pm 0.21_{B\widetilde{B}_S}
    \; \pm 1.39_{1/m_b} \pm 0.09_{\textrm{input}}
    \right) \times
    10^{-2}~\mbox{ps}^{-1}\ 
     (\overline{\textrm{MS}})\,,
   \nonumber\\
  \Delta \Gamma_s 
  ~=&~ \left(
      {{7.77^{+0.59}_{-0.62}}_{\textrm{scale}}}
      {{}^{+0.16}_{-0.34}}_{\textrm{scale, $1/m_b$}}
    %{\pm0.25}_{\textrm{scale, $1/m_b$}} 
      \pm 0.20_{B\widetilde{B}_S}
    \; \pm 1.39_{1/m_b} \pm 0.09_{\textrm{input}}
    \right) \times
    10^{-2}~\mbox{ps}^{-1}\ 
                      (\textrm{PS})\,.
          \label{eq::dGamma_s}
\end{align}

The results in the $\overline{\rm MS}$ and PS schemes are averaged in our final result. For the uncertainty, we add the upper and lower bounds in quadrature, which are then symmetrised and averaged across the schemes. We obtain
\begin{equation}
  \Delta\Gamma_s = {{(0.078 \pm 0.015)}} ~\mbox{ps}^{-1}\,, \label{eq:dGs_average}
\end{equation}
which is in excellent agreement with the experimental value \cite{HFLAV:2024ctg}
\begin{equation}
    \Delta \Gamma_s^\text{exp} = \SI{0.0781(35)}{\per\pico\second}\,, \label{eq:dgsexp}
\end{equation}

\subsubsection{$B_d$ system}

In the $B_d$ system, our predictions for $\Delta\Gamma_d/\Delta M_d$ are given by
\begin{align}
  \frac{\Delta \Gamma_d}{\Delta M_d}
  ~=&~ \left(
      {{3.83^{+0.49}_{-0.53}}_{\textrm{scale}}  }
      {{}^{+0.12}_{-0.20}}_{\textrm{scale, $1/m_b$}}
%    {\pm0.16}_{\textrm{scale, $1/m_b$}} 
      \pm 0.11_{B\widetilde{B}_S}
    \pm 0.79_{1/m_b} \pm 0.06_{\textrm{input}}
    \right) \times
    10^{-3}\ (\textrm{pole})\,, \nonumber\\
      \frac{\Delta \Gamma_d}{\Delta M_d} 
  ~=&~ \left(
      {{4.29^{+0.19}_{ -0.40}}_{\textrm{scale}}   }
      {{}^{+0.12}_{-0.20}}_{\textrm{scale, $1/m_b$}}
    %{\pm0.16}_{\textrm{scale, $1/m_b$}} 
      \pm 0.12_{B\widetilde{B}_S}
    \pm 0.79_{1/m_b} \pm 0.05_{\textrm{input}}
   \right) \times 10^{-3}\ (\overline{\textrm{MS}})\,,
   \nonumber\\
  \frac{\Delta \Gamma_d}{\Delta M_d} ~=&~ 
    \left( 
    {{ 4.22^{+0.33}_{ -0.35}}_{\textrm{scale}} } 
      {{}^{+0.12}_{-0.20}}_{\textrm{scale, $1/m_b$}}
    %{\pm0.16}_{\textrm{scale, $1/m_b$}} 
      \pm 0.12_{B\widetilde{B}_S}
      \pm 0.79_{1/m_b}
          \pm 0.05_{\textrm{input}}
    \right) \times 10^{-3}\ (\textrm{PS})\,.
          \label{eq::dGdM_d}
\end{align}

\begin{figure}[t]
  \begin{center}
      \includegraphics[width=0.8\textwidth]{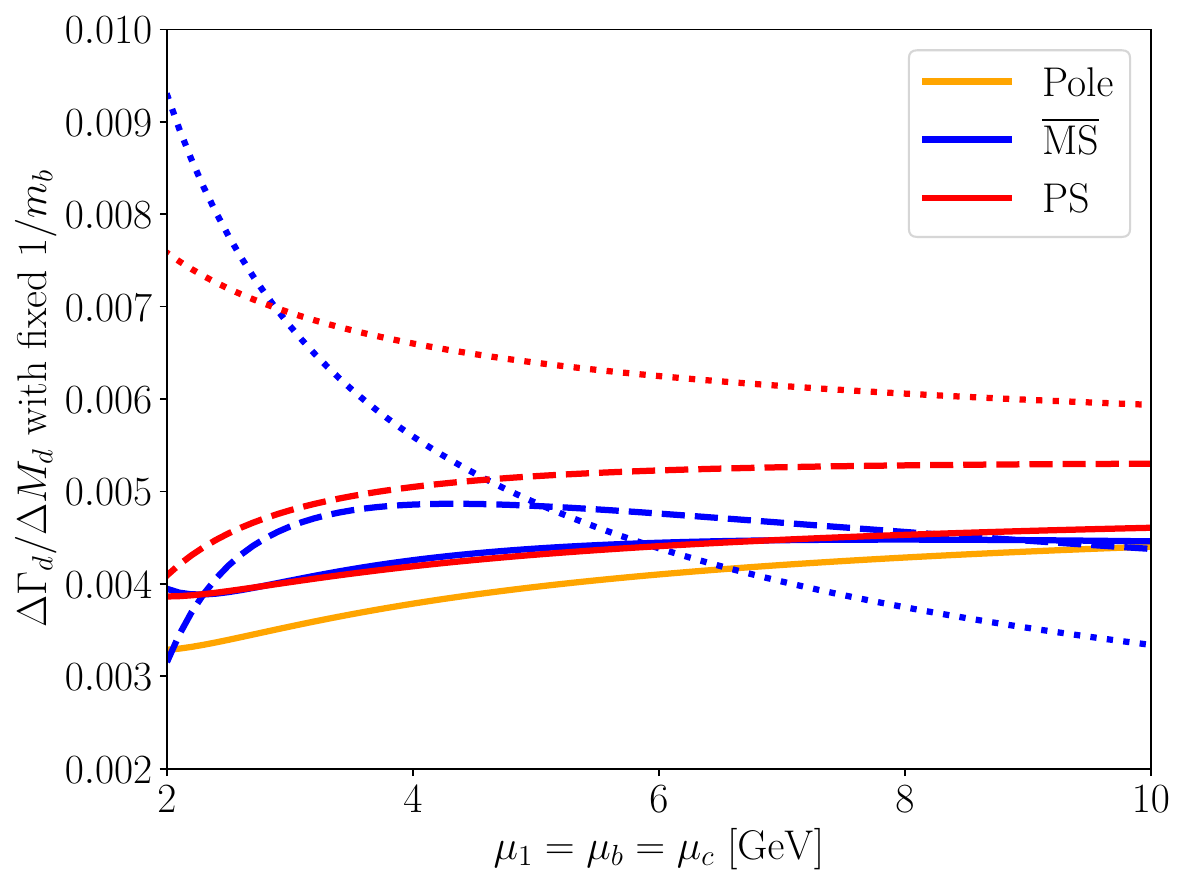}
  \end{center}    \caption{\label{fig::DelGam_mu1_Bd} 
      Renormalisation scale dependence at LO (short dashes),
      NLO (long dashes) and NNLO (solid) for $\Delta\Gamma_d/\Delta M_d$. The scale variation shown is of the leading-power terms where the scales $\mu_1=\mu_b=\mu_c$ are varied together.
      }  
%  ~\\[-5mm]
%\hrule    
\end{figure}

As for the $B_s$ system, we can compare with Ref.~\cite{Gerlach:2025tcx} to show the impact of our updated calculation with a deeper expansion in $z=m_c^2/m_b^2$ at NLO for the penguin contributions and the penguin operators at NNLO. The NLO penguin contributions 
including all terms amount to about $26\%$ of the two leading terms in the $\overline{\text{MS}}$ scheme, which is similar to the $B_s$ case. At NNLO, the penguin operators contribute about $12\%$ to the total NNLO part and their inclusion brings down the scale uncertainty of the leading $1/m_b$ terms by almost 10\%. The overall result is changed by a similar amount to the $B_s$ system with the central value shifting by about $2\%$.

In Ref.~\cite{Beneke:2003az} it was found that at NLO the ratio $\Delta\Gamma_d/\Delta M_d$ receives only small corrections of $2\%$ from the terms proportional to $\lambda_u^d/\lambda_t^d$. i.e.~the second and third terms in Eq.~\eqref{eq::Gam12}. At NNLO the same behaviour can be observed with the $\lambda_u^d/\lambda_t^d$ terms contributing in total only $1.9\%$. As the ratio $\Delta\Gamma_s/\Delta M_s$ also receives negligible corrections from the terms proportional to the CKM elements, the two ratios have very similar numerical values, see Eqs.~(\ref{eq::dGdM}) and~(\ref{eq::dGdM_d}).

Using the experimental value~\cite{HFLAV:2024ctg}
\begin{equation}
  \Delta M_d^{\rm exp} = (0.5069 \pm 0.0019)\, \mbox{ps}^{-1}\,,
  \label{eq:dMd_exp}
\end{equation}
we obtain 
\begin{align}
  \Delta \Gamma_d
  ~=&~ \left(
      {{1.94^{+0.25}_{-0.27}}_{\textrm{scale}}    }
      {{}^{+0.06}_{-0.10}}_{\textrm{scale, $1/m_b$}}
    %{\pm0.08}_{\textrm{scale, $1/m_b$}} 
      \pm 0.06_{B\widetilde{B}_S}
    \; \pm 0.40_{1/m_b} \pm 0.03_{\textrm{input}}  
    \right) \times
    10^{-3}\mbox{ps}^{-1}\ 
      (\textrm{pole})\,, \nonumber\\
      \Delta \Gamma_d
  ~=&~ \left(
      {{2.17^{+0.10}_{-0.20}}_{\textrm{scale}}  }
      {{}^{+0.06}_{-0.10}}_{\textrm{scale, $1/m_b$}}
    %{\pm0.08}_{\textrm{scale, $1/m_b$}} 
      \pm 0.06_{B\widetilde{B}_S}
    \; \pm 0.40_{1/m_b} \pm 0.03_{\textrm{input}}  
    \right) \times
    10^{-3}\mbox{ps}^{-1}\ 
     (\overline{\textrm{MS}})\,,
   \nonumber\\
  \Delta \Gamma_d
  ~=&~ \left(
      {{2.14^{+0.17}_{-0.18}}_{\textrm{scale}}   }
      {{}^{+0.06}_{-0.10}}_{\textrm{scale, $1/m_b$}}
    %{\pm0.08}_{\textrm{scale, $1/m_b$}} 
      \pm 0.06_{B\widetilde{B}_S}
    \; \pm 0.40_{1/m_b} \pm 0.03_{\textrm{input}}  
    \right) \times
    10^{-3} \mbox{ps}^{-1}\ 
                      (\textrm{PS})\,
          \label{eq::dGamma_d_2}
\end{align}
from the results in Eq.~(\ref{eq::dGdM_d})
after multiplication with $\Delta M_d^{\rm exp}$.
Adding the uncertainties in quadrature for the upper and lower bounds separately, symmetrising the total uncertainty in each scheme and averaging the results for the $\overline{\text{MS}}$ and PS schemes we obtain
\begin{equation}
    \Delta \Gamma_d = {{(0.00215 \pm 0.00045)}} ~\mbox{ps}^{-1}\,. \label{eq:res_dGd_first}
\end{equation}

\subsection{$a_{\rm fs}^q$}

From the calculation performed
for $\Delta\Gamma_q/\Delta M_q$ we can extract results for $a_{\rm fs}$ by taking the imaginary part of $\Gamma_{12}^q/M_{12}^q$. For the $B_s$ system we obtain\footnote{
The uncertainties from the $1/m_b$ bag parameters are smaller than those given in Ref.~\cite{Gerlach:2025tcx} since the correlation
between $R_3$, $\tilde{R}_3$ and $\tilde{R}_2$ has been accounted for in Eqs.~(\ref{eq:afsnum_s}) and~(\ref{eq:afsnum_s}).}
\begin{align*}
  a_{\rm fs}^s
  &= 
      \left(
      {{2.27^{{+0.00}}_{-0.03}}_{\textrm{scale}}} 
      { {{}^{+0.01}_{-0.00}}_{\textrm{scale, $1/m_b$}}}
    %{\pm0.01}_{\textrm{scale, $1/m_b$}} 
      {\pm 0.01_{B\widetilde{B}_S}}
      {\pm 0.04_{1/m_b}}
      {\pm 0.07_{\textrm{input}}}
      \right) \times
      10^{-5}\ (\textrm{pole})\,,\\
  a_{\rm fs}^s
  &= 
      \left(
      {{2.24^{{+0.10}}_{-0.18}}_{\textrm{scale}}}
      { {{}^{+0.01}_{-0.00}}_{\textrm{scale, $1/m_b$}}}
    %{\pm0.01}_{\textrm{scale, $1/m_b$}} 
      {\pm 0.01_{B\widetilde{B}_S}}
      {\pm 0.04_{1/m_b}}
      {\pm 0.07_{\textrm{input}}}
      \right) \times
      10^{-5}\ (\overline{\rm MS})\,,\\
  a_{\rm fs}^s
  &= 
      \left(
      {{2.30^{{+0.03}}_{-0.07}}_{\textrm{scale}}}
      { {{}^{+0.01}_{-0.00}}_{\textrm{scale, $1/m_b$}}}
    %{\pm0.01}_{\textrm{scale, $1/m_b$}} 
      {\pm 0.01_{B\widetilde{B}_S}}
      {\pm { 0.04_{1/m_b}} }
      {\pm 0.07_{\textrm{input}}}
      \right) \times
      10^{-5}\ (\textrm{PS})\,.\numberthis
  \label{eq:afsnum_s}
\end{align*}
For the $B_d$ system we have\footnote{Note that there is a typo in Eq.~(96) of Ref.~\cite{Gerlach:2025tcx}: The lower limit of the ``scale, $1/m_b$'' uncertainty should read ``$-0.01$'' and not ``$-0.08$''.}
\begin{align*}
  a_{\rm fs}^d
  &= -
      \left(
      {{5.18^{{+0.00}}_{{-0.08}}}_{\textrm{scale}}   }
      {{}^{{+0.03}}_{{-0.01}}}_{\textrm{scale, $1/m_b$}}
    %{\pm0.06}_{\textrm{scale, $1/m_b$}} 
      \pm {0.03_{B\widetilde{B}_S}}
      \pm {0.09_{1/m_b} }
      \pm 0.16_{\textrm{input}}
      \right) \times
      10^{-4}\ (\textrm{pole})\,,\\
  a_{\rm fs}^d
  &= -
      \left(
      {{5.12^{{+0.23}}_{{-0.41}}}_{\textrm{scale}}  }
      {{}^{{+0.03}}_{{-0.01}}}_{\textrm{scale, $1/m_b$}}
    %{\pm0.06}_{\textrm{scale, $1/m_b$}} 
      \pm {0.03_{B\widetilde{B}_S}  }
      \pm {0.09_{1/m_b}   }
      \pm {0.16_{\textrm{input}}}
      \right) \times
      10^{-4}\ (\overline{\rm MS})\,,\\
  a_{\rm fs}^d
  &= -
      \left(
      {{5.26^{{+0.07}}_{{-0.15}}}_{\textrm{scale}}   }
      {{}^{{+0.03}}_{{-0.01}}}_{\textrm{scale, $1/m_b$}}
    %{\pm0.06}_{\textrm{scale, $1/m_b$}} 
      \pm {0.03_{B\widetilde{B}_S}}
      \pm { 0.09_{1/m_b} }
      \pm 0.16_{\textrm{input}}
      \right) \times
      10^{-4}\ (\textrm{PS})\,.\numberthis
  \label{eq:afsnum_d}
\end{align*}
Comparing our updated results with Ref.~\cite{Gerlach:2025tcx}, we note that the central values or both $B_s$ and $B_d$ are shifted by $0.5\%$ in the $\overline{\textrm{MS}}$ scheme. In the $B_s$ system, the expansion of the NLO penguin contributions beyond $z^1$ is as big as this leading term but has the opposite sign. Similarly, both for $B_s$ and $B_d$ the NNLO penguin terms make up 
about $1\%$ of the NNLO contribution.
\begin{figure}[t]
  \begin{center}
    \begin{tabular}{c}
      \includegraphics[width=0.67\textwidth]{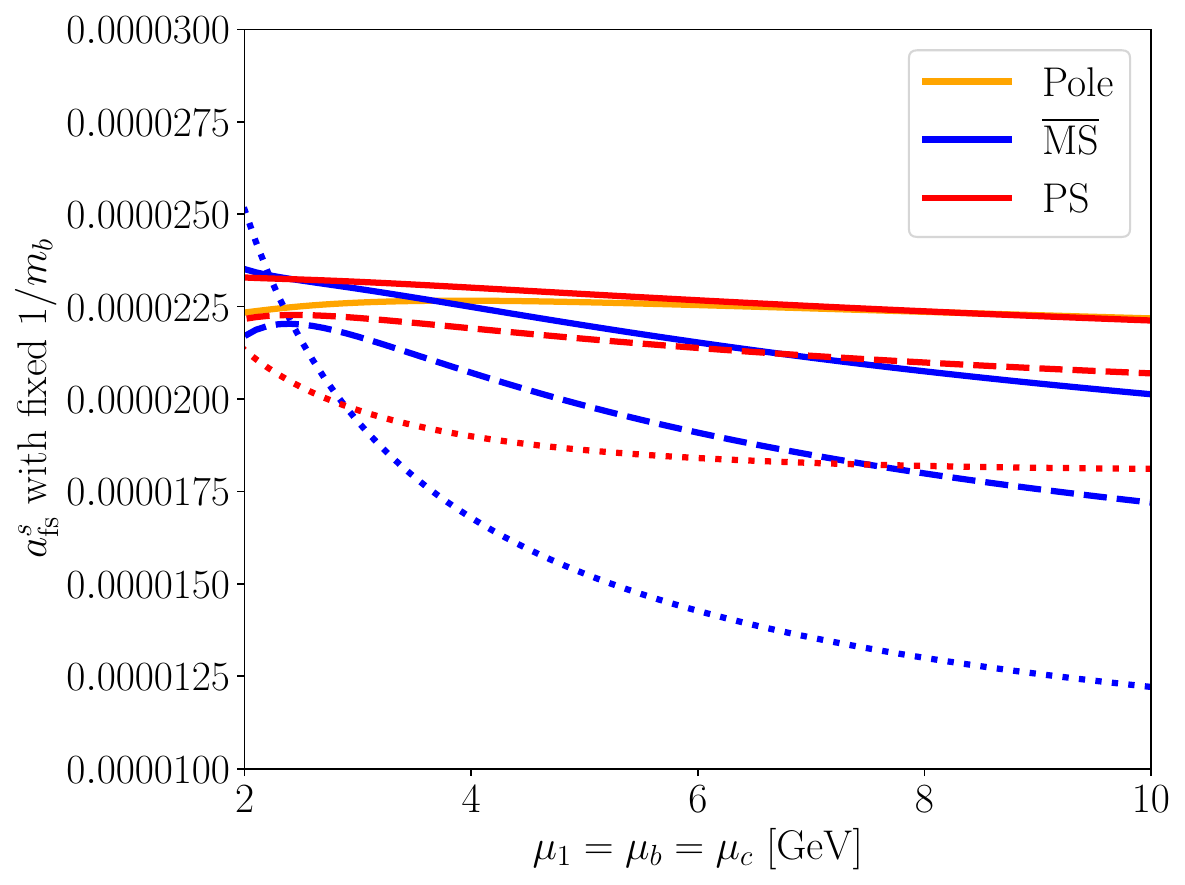}
    \\ 
    %(a) \\
      \includegraphics[width=0.67\textwidth]{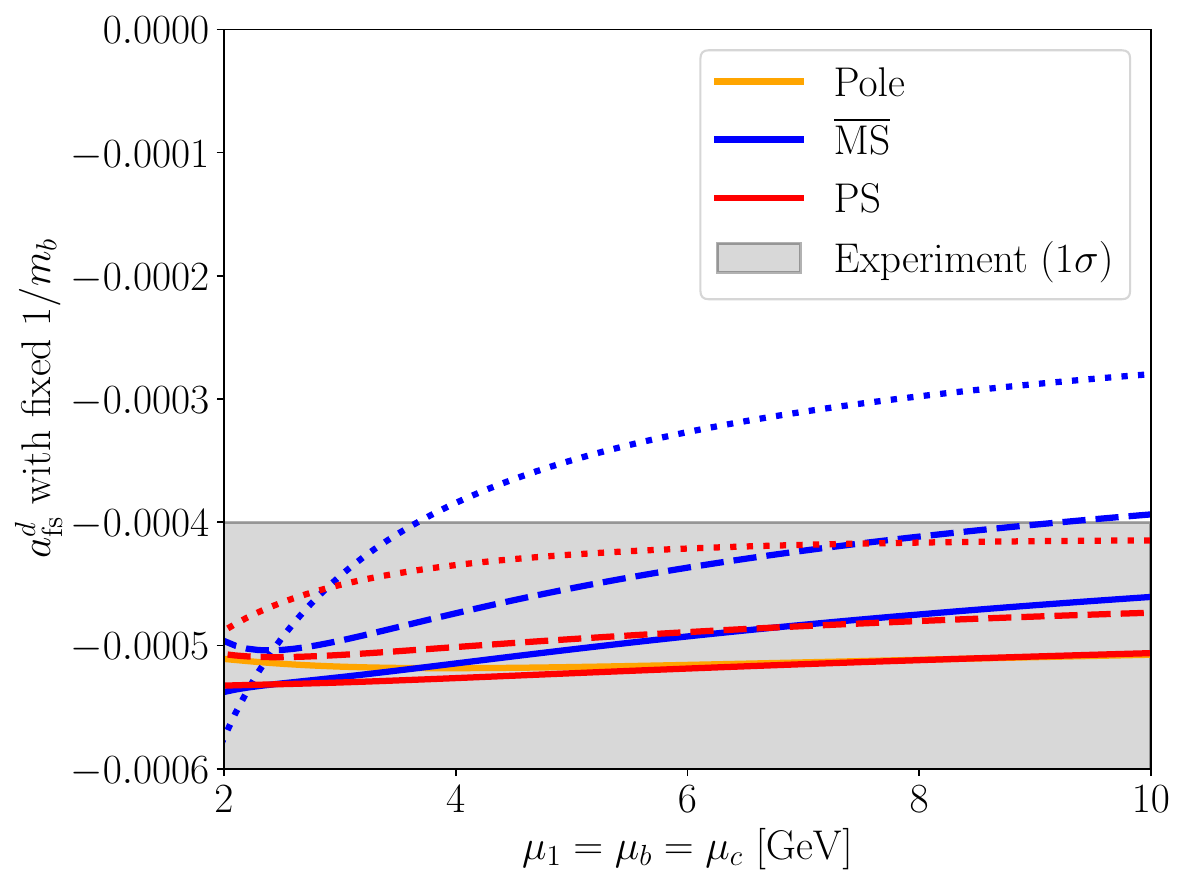}
    \\ 
    %(b)
    \end{tabular}
  \end{center}   
  \vspace*{-1.5em}
  \caption{\label{fig::afd_afs_mu1}
  Renormalisation scale dependence at LO (short dashes), NLO (long dashes) and
      NNLO (solid) for $a_{\rm fs}^s$ (top) and $a_{\rm fs}^d$ (bottom). The scales $\mu_1=\mu_b=\mu_c$ are varied simultaneously for the leading-power terms while the $1/m_b$-suppressed terms are kept fixed at the central scale.}
%~\\[-5mm]
%\hrule
\end{figure}

Adding the uncertainties in quadrature for the upper and lower bounds, symmetrising the total uncertainty in each scheme, and averaging the results for the $\overline{\text{MS}}$ and PS schemes we obtain
\begin{align*}
    a_{\rm fs}^s &= \left({2.27 \pm 0.13} \right) \times 10^{-5}\,,\\
    a_{\rm fs}^d &= -\left({5.19 \pm 0.30} \right) \times 10^{-4}\,.\numberthis
\end{align*}

\subsection[The double ratio 
$({(\Delta\Gamma_d /\Delta M_d) / (\Delta \Gamma_s/ \Delta M_s)}$]
{The double ratio 
$\boldsymbol{(\Delta\Gamma_d /\Delta M_d) / (\Delta \Gamma_s/ \Delta M_s) }$}

The double ratio
\begin{equation}
    r_{ds} \equiv \frac{\Delta\Gamma_d}{\Delta M_d} \times \frac{\Delta M_s}{\Delta \Gamma_s}
\end{equation}
is phenomenologically significant as it can be used to obtain a very accurate prediction for $\Delta \Gamma_d$. This is because the ratio $r_{ds}$ is very close to one, theoretical uncertainties cancel, and the other three quantities, $\Delta M_d$,
$\Delta M_s$ and $\Delta \Gamma_s$ are measured to high precision. 

The ratio happens to be $\approx 1$ since most of the dependence on the hadronic matrix elements cancels in $\Gamma_{12}/M_{12}$, and additionally the difference stemming from the CKM elements is small. Parametrising the ratio for each of the $B_q$ systems as \cite{Beneke:2003az}
\begin{align}
 \frac{\Gamma_{12}^q}{M_{12}^q} &=\, 10^{-4} \lt[ c^q + a^q \frac{\lambda_u^q}{\lambda_t^q} + b^q \frac{\left(\lambda_u^{q}\right)^2}{\left(\lambda_t^{q}\right)^2}\rt]\,,
 \label{eq:gabc}
\end{align}
we find that the constants $a$, $b$ and $c$ only have a weak dependence on the ratio of the hadronic matrix elements $\langle Q \rangle$ and $\langle \widetilde{Q}_S \rangle$. In the $B_s$ system, the ratio $\lambda_u^s/\lambda_t^s$ is small, and for the $B_d$ system the linear term for $\Delta\Gamma_d/\Delta M_d$
is proportional to $\cos\alpha$, which is close to zero near the apex of the UT (see \fig{fig:ut}), while the quadratic correction is doubly Cabibbo-suppressed~\cite{Beneke:2003az}. As mentioned above, we find that the sum of the $a^q$ and $b^q$ terms contributes about $0.2\%$ and $1.9\%$ for $B_s$ and $B_d$, respectively.

One additional attractive feature of the double ratio $r_{ds}$ is that the uncertainty  of the ratio of the hadronic matrix elements for $B_d$ and $B_s$ is substantially smaller than that of the individual matrix elements. This feature stems from the fact that the matrix element are equal in the limit of exact SU(3)$_{\rm F}$ symmetry. Moreover, the SU(3)$_{\rm F}$ breaking in the decay constant is precisely known and furthermore SU(3)$_{\rm F}$ breaking in the bag parameters is found to be small where calculated.
Here, we use the parametrisation of the leading order matrix elements
\begin{equation}
    \xi^2 \equiv \frac{f_{B_s}^2 B_{B_s}}{f_{B_d}^2 B_{B_d}}\,, \label{eq:xi_def}
\end{equation}
which has been determined very precisely in Ref.~\cite{Dowdall:2019bea},
see Tab~\ref{tab::input}. For the bag parameters for $\langle \widetilde{Q}_S\rangle$, we calculate a similar ratio
\begin{equation}
    \xi_S^2 \equiv \frac{f_{B_s}^2 \widetilde{B}^\prime_{S,B_s} }{f_{B_d}^2 \widetilde{B}^\prime_{S,B_d}}  \label{eq:xis_def}
\end{equation}
using the numerical value from Ref.~\cite{Dowdall:2019bea} (see Tab.~XI),
\begin{alignat}{3}
    \frac{B^{(3)}_{B_s}}{B^{(3)}_{B_d}} &= \SI{1.092(34)}{}\,,%\qquad && \eta_3^s = \SI{0.534(12)}{}\,,\qquad && \eta_3^d = \SI{0.536(12)}{}\,,
    \label{eq:B3_ratio}
\end{alignat}
and the ratio of the parameters $\eta_3^s$ and $\eta_3^d$, which we calculate from the definition given in Ref.~\cite{Dowdall:2019bea},
\begin{equation}
    \frac{\eta_3^{s}}{\eta_3^{d}} = \SI{0.99651(39)}{}\,.
    \label{eq:eta_ratio}
\end{equation}
For the bag parameters in our convention
\begin{equation}
    \widetilde{B}'_{S} \equiv \eta_3 B^{(3)}\,,
\end{equation}
we combine the results from Eqs.~\eqref{eq:B3_ratio} and \eqref{eq:eta_ratio} to obtain
\begin{equation}
    \frac{\widetilde{B}'_{S,B_s}}{\widetilde{B}'_{S,B_s}} = \SI{1.088(34)}{}\,.
    %\frac{\widetilde{B}'_{S,B_s}}{\widetilde{B}'_{S,B_s}} = \SI{1.088(48)}{}\,.
\end{equation}
With the ratio of the decay constants from Tab.~\ref{tab::input} 
we therefore obtain
\begin{equation}
    \xi_S = \SI{1.263(20)}\,.
\end{equation}

The bag parameters for the 
$1/m_b$ suppressed matrix elements
can also be found in the literature~\cite{Dowdall:2019bea,Davies:2019gnp}.
However, using the results for the $B_s$ and $B_d$ mesons in the double ratio would vastly overestimate the uncertainty. For this reason
we proceed as follows: We adopt the  values for the 
bag parameters of the $B_s$ meson from the literature~\cite{Dowdall:2019bea,Davies:2019gnp}. Afterwards we add manually
SU(3)$_\text{F}$-breaking effects. To investigate this effect, we take the numerical values of the bag parameters for $B_s$ mesons obtained from lattice calculations as quoted in Ref.~\cite{Gerlach:2025tcx} and parametrise the corresponding values in the $B_d$ system using
%\begin{equation}
%   \zeta  \equiv  \frac{B^{(s)}_i}{ B^{(d)}_i}\,.\label{eq:zeta_su3f}
%\end{equation}
\begin{align}
\bra{B_q} R_i \ket{\bar{B}_q} &\equiv f^2_{B_q} M^2_{B_q} B_{R_i, B_q} \,, & B_{R_i,B_d} &\equiv \frac{B_{R_i, B_s} }{\zeta}\,.\label{eq:zeta_su3f}
\end{align}
The variable $\zeta$ encodes the extent of SU(3)$_{\rm F}$ symmetry breaking in both the bag parameters. For our numerical analysis we choose
\begin{equation}
    \zeta = 1.0 \pm 0.1\,,
\end{equation}
which allows for up to $10\%$ SU(3)$_\text{F}$ breaking. We note that with this parametrisation, only the ratio of the decay constants as given in Tab.~\ref{tab::input} is required as numerical input for $r_{ds}$, not the individual decay constants.

\begin{figure}[tb]
  \begin{center}
      \includegraphics[width=0.8\textwidth]{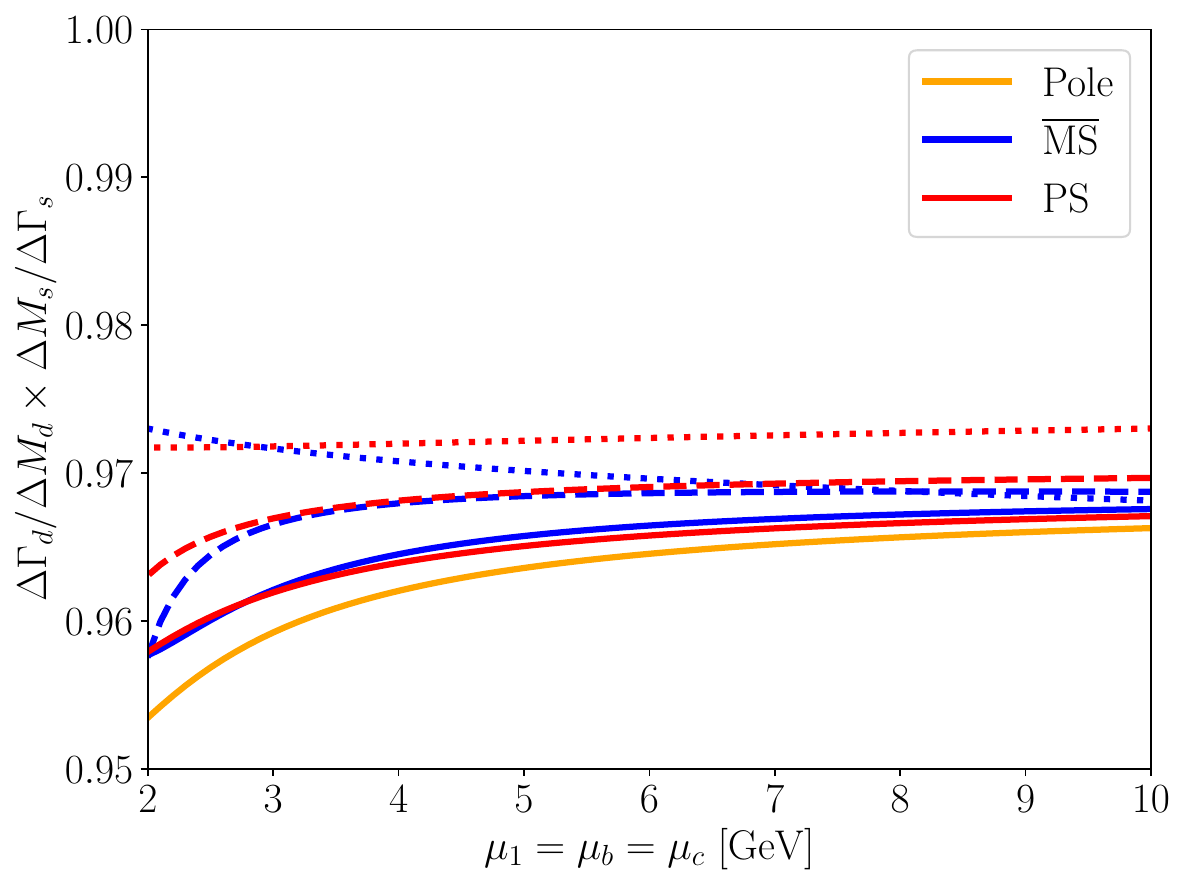}
  \end{center}    \caption{\label{fig:double_ratio_scale} 
      Renormalisation scale dependence at LO (short dashes),
      NLO (long dashes) and NNLO (solid) for $r_{ds}$. The scale variation shown is of the leading and sub-leading terms in the HQE where the scales $\mu_1=\mu_b=\mu_c$ are varied together. Note that for the NLO and NNLO scale dependence, the respective $\alpha_s$ corrections were included in the individual $\Gamma_{12}/M_{12}$ ratios without truncating the double ratio in $\alpha_s$.
      }  
%  ~\\[-5mm]
%\hrule    
\end{figure}

Our results for the double ratio read
\begin{align}
    r_{ds} &= 
      {{0.9624^{+0.0034}_{-0.0082}}_{\textrm{scale, comb.}}
%      {{{}^{+0.xx}_{-0.xx}}_{\textrm{scale, $1/m_b$}} } 
      {\pm 0.012_{B\widetilde{B}_S}}
      {\pm 0.040_{1/m_b}} { \pm 0.003_{\textrm{input}}}} \ (\textrm{pole})\,, \nonumber\\
    r_{ds} &= 
      {{0.9648^{{ +0.0025}}_{{ -0.0067}}}_{\textrm{scale, comb.}}
%      {{{}^{+0.xx}_{-0.xx}}_{\textrm{scale, $1/m_b$}} } 
      {\pm 0.011_{B\widetilde{B}_S}}
      {\pm 0.036_{1/m_b}} { \pm 0.003_{\textrm{input}}}} \ (\overline{\textrm{MS}})\,, \nonumber\\
    r_{ds} &= 
      {{0.9642^{+0.0025}_{-0.0058}}_{\textrm{scale, comb.}}
%      {{{}^{+0.xx}_{-0.xx}}_{\textrm{scale, $1/m_b$}} } 
      {\pm 0.012_{B\widetilde{B}_S}}
      {\pm 0.036_{1/m_b}} { \pm 0.003_{\textrm{input}}}}\ (\textrm{PS})\,. 
\end{align}
Note that the uncertainty of the $1/m_b$-suppressed terms stems mainly from the unknown size of SU(3)$_\text{F}$ breaking. Neglecting the uncertainty of $\zeta$, the uncertainty labelled $1/m_b$ is just $\pm0.008$.
In Fig.~\ref{fig:double_ratio_scale} we show the remaining renormalisation scale dependence of the leading and sub-leading $1/m_b$ terms combined of the double ratio. It is worth commenting on the increasing scale dependence from LO to NNLO. This is due to the fact that the scale dependence of the $\Gamma_{12}/M_{12}$ ratios in the $B_s$ and $B_d$ systems are almost identical, see Figs.~\ref{fig::DelGam_mu1} and \ref{fig::DelGam_mu1_Bd}. Therefore, small deviations which appear at higher orders lead to a seemingly worse convergence of the perturbative series. However, overall the scale dependence of the leading power terms is still small and the prediction of $r_{ds}$ very accurate.

Averaging the results of the $\overline{\text{MS}}$ and PS schemes and adding the uncertainties of the upper and lower bounds in quadrature before symmetrising, we obtain our final result for the double ratio
\begin{equation}
    r_{ds} = 0.965 \pm 0.038\,.
\end{equation}
Using this together with the experimental values of the mass differences from \eq{eq:dMd_exp} and \eq{eq:dMs_exp} as well as $\dg_s$ from \eq{eq:dgsexp}
we obtain our most accurate theoretical prediction of 
\begin{equation}
    \Delta \Gamma_d = {{(0.00215 \pm 0.00013)}} ~\mbox{ps}^{-1}\,.
\end{equation}
Comparing this result to the previously obtained value in \eq{eq:res_dGd_first}, the uncertainty has been reduced by $70\%$ thanks to the better control of the hadronic uncertainties in the double ratio $r_{ds}$.

\subsection[Unitarity triangle constraints from ${a_{\rm fs}^d}$ and ${\dg_d/\dg_s}$]{Unitarity triangle constraints from $\boldsymbol{a_{\rm fs}^d}$ and $\boldsymbol{\dg_d/\dg_s}$}
As shown in Fig.~\ref{fig:ut}, future measurements of $a_{\rm fs}^d$ and $\dg_d$
will lead to important constraints on the apex $(\bar\rho,\bar\eta)$ of the unitarity triangle.

We update the constraint from $a_{\rm fs}^d$ shown in Fig.~8 of Ref.~\cite{Gerlach:2025tcx} and further discuss the ratio $\dg_d/\dg_s$, from which many of the theoretical uncertainties cancel. For these analyses, it is useful to define the quantities $a$, $b$, and $c$ via \eq{eq:gabc}. With \eq{eq::Gam12} one finds
\begin{align}
c^q &=\; - 10^{4} \, \frac{(\lambda_t^{q})^2}{M_{12}^q}\,  \Gamma_{12}^{cc}\,, \nonumber\\
 a^q &=\,  - 2\cdot 10^{4} \, \frac{(\lambda_t^{q})^2}{M_{12}^q} \,  
       \left(\Gamma_{12}^{cc}-\Gamma_{12}^{uc}\right), \nn 
 b^q &=\,  - 10^{4} \, \frac{(\lambda_t^{q})^2}{M_{12}^q} \,       
      \left(\Gamma_{12}^{uu}+\Gamma_{12}^{cc}-2\Gamma_{12}^{uc}\right),
\label{eq:abc}
\end{align}
which is convenient because $\left(\lambda_t^{q}\right)^2/M_{12}^q$ and thus $a$, $b$, and $c$ are real, and the ratios  $\lambda_u^q/\lambda_t^q$ in \eq{eq:gabc} have simple expressions in terms of the improved Wolfenstein parameters. For $q=d$ the latter are given 
in \eq{eq:utp} and for $q=s$ we have
\begin{align}
  \frac{\lambda_u^s}{\lambda_t^s} &=\, \lambda^2 (-\bar\rho + i \bar\eta) 
  \, +\, \lambda^4 \lt( -\bar\rho (1-\bar\rho) -\bar\eta^2+ i \bar\eta (1-2\bar\rho) \rt)
  \, +\, {\cal O} (\lambda^6), \label{eq:utps}
\end{align}
where $\lambda \simeq |V_{us}|$ with  $\lambda =0.225$ is a 
Wolfenstein parameter.

The ratio of the width differences can be written as
\begin{align}
    \frac{\dg_d}{\dg_s} &=\, \frac{1}{\xi^2}\,\frac{M_{B_d}}{M_{B_s}}\, \left\lvert\frac{\lambda_t^d}{\lambda_t^s}\right\rvert^2 \,
   \frac{\ds c^d+  a^d \, \real \frac{\lambda_u^d}{\lambda_t^d} + 
   b^d \, \real \frac{(\lambda_u^{d})^2}{(\lambda_t^{d})^2}}{
   \ds c^s +  a^s\, \real \frac{\lambda_u^s}{\lambda_t^s} + 
   b^s \,\real \frac{(\lambda_u^{s})^2}{(\lambda_t^{s})^2}
   } \,, \label{eq:drat_ds}
\end{align}
where the ratio $\xi^2$ from Eq.~\eqref{eq:xi_def} was used. For Eq.~\eqref{eq:drat_ds} we need a further parametrisation of the CKM factors in terms of $\bar\rho$ and $\bar\eta$,
\begin{equation}
\left\lvert\frac{\lambda_t^d}{\lambda_t^s}\right\rvert^2 =\lambda^2 \left(\bar\eta^2 + (1-\bar\rho)^2 \right) \times \left(1 + \lambda^2 (1-2\bar{\rho})\right) + \mathcal{O}\left(\lambda^6\right)\,.\label{eq:lambdat_ratio_param}
\end{equation}

The ratio in Eq.~(\ref{eq:drat_ds}) can be further simplified since the numerical values of $\{a^d,b^d,c^d\}$ and $\{a^s,b^s,c^s\}$ agree very well within their (perturbative) 
uncertainties, see Tabs.~\ref{tab:afs_d} and~\ref{tab:afs_s}. We can thus simplify the ratio in Eq.~\eqref{eq:drat_ds} to
\begin{align}
    \frac{\dg_d}{\dg_s} &=\,  \frac{1}{\xi^2}\,\frac{M_{B_d}}{M_{B_s}} \, \left\lvert\frac{\lambda_t^d}{\lambda_t^s}\right\rvert^2 \,
   \frac{\ds 1+  \frac{a}{c} \real \frac{\lambda_u^d}{\lambda_t^d} + 
   \frac{b}{c}\real \frac{(\lambda_u^{d})^2}{(\lambda_t^{d})^2}}{
   \ds 1+  \frac{a}{c}\real \frac{\lambda_u^s}{\lambda_t^s} + 
   \frac{b}{c}\real \frac{(\lambda_u^{s})^2}{(\lambda_t^{s})^2}
   } \,, \label{eq:drat}
\end{align}
where the superscripts were dropped. %It should be noted here that we do not account for the correlation of the bag parameters as shown in Eq.~(\ref{eq:zeta_su3f}) and therefore we slightly overestimate the input uncertainty from the bag parameters. However, the overall input uncertainty is still reasonably small to demonstrate that $\Delta\Gamma_d/\Delta\Gamma_s$ is a good observable for constraining the apex of the CKM triangle, see Figs.~\ref{fig:double_constraints_CKM} and \ref{fig:double_constraints_CKM_all}.

The flavour-specific CP asymmetry $a_{\rm fs}^d$ is similarly parametrised as
\begin{equation}
  a_{\rm fs}^d = \left[a\, \text{Im} \frac{\lambda_u^d}{\lambda_t^d} + b\, \text{Im} \frac{(\lambda_u^d)^2}{(\lambda_t^d)^2} \right]\times 10^{-4}
  \,.
  \label{eq::afs_pred}
\end{equation}

By inserting \eqsand{eq:utp}{eq:utps} into \eq{eq:drat} one 
finds the constraint on $(\bar\rho,\bar\eta)$ defined by $\dg_d/\dg_s$. Analogously the constraint from $a_\text{fs}^d$ is obtained by inserting Eq.~\eqref{eq:utp} into Eq.~\eqref{eq::afs_pred}.

In order to construct the constraints on the CKM triangle, we calculate the quantities $a$, $b$ and $c$ in Eq.~\eqref{eq:gabc} to LO, NLO and NNLO. These constants can be further decomposed into a leading and sub-leading $1/m_b$-suppressed contribution,
\begin{align*}
    a &= a_0 + a_1\,,\\
    b &= b_0 + b_1\,,\\
    c &= c_0 + c_1\,, \numberthis
\end{align*}
where the subscripts $0$ and $1$ denote the leading and sub-leading terms of the HQE respectively. The sub-leading contributions of $a^d$ and $b^d$ have been calculated previously and are given in Ref.~\cite{Gerlach:2025tcx}. For completeness we list them again here together with the new value for $c$,
\begin{align}
    a^d_1 &= {0.622^{+0.073}_{-0.020}}_\textrm{scale} \pm {0.35_{\textrm{para}}}\,,\nonumber\\
    b^d_1 &= {0.091^{+0.011}_{-0.003}}_\textrm{scale} \pm {0.031_{\textrm{para}}}\,,\nonumber\\
    {c^d_1} &= {{15.36^{+1.98}_{-1.22}}_\textrm{scale} \pm {8.03_{\textrm{para}}}\,,}
\end{align}
where the uncertainties from varying all input parameters have been attributed to the sub-leading terms, denoted ``para''. The leading contributions $a_0, b_0, c_0$ hence only have a scale uncertainty. The ``para'' uncertainty is dominated by $m_c$ for $a_1$ and $b_1$ while 
$\widetilde{r}_2^q$ has the biggest impact for $c_1$. Note that the scale uncertainties are obtained as before by varying $\mu_1=\mu_b=\mu_c$ simultaneously between $\SI{2.1}{GeV}$ and $\SI{8.4}{GeV}$. The sub-leading term is only known to LO and is therefore the same for all renormalisation schemes up to minor differences in the parametric uncertainties. For the $B_s$ system we find similarly
\begin{align}
    a^s_1 &= {{0.616^{+0.074}_{-0.025}}_\textrm{scale} \pm { 0.35_{\textrm{para}}} }\,,\nonumber\\
    b^s_1 &= {{0.090^{+0.011}_{-0.004}}_\textrm{scale} \pm { 0.030_{\textrm{para}}} }\,,\nonumber\\
    c^s_1 &= {{15.28^{+1.90}_{-0.93}}_\textrm{scale} \pm { 7.90_{\textrm{para}}}\,,}
\end{align}

For the leading-$1/m_b$ contributions, the results in the $\overline{\textrm{MS}}$ and PS schemes are given in Tab.~\ref{tab:afs_d}. When comparing to the results obtained for the current-current contributions only at NNLO in Ref.~\cite{Gerlach:2025tcx}, we note that $b_0$ is unchanged because the penguin operators do not yield additional contributions to the $(\lambda_u^d)^2$ term. In fact, only the mixed current-current penguin contributions affect $a_\textrm{fs}$, and only in terms of $a_0$. The results for $c_0$ are entirely new.
\begin{table}[t]
    \centering
    \begin{tabular}{cc}
    \begin{tabular}{@{} l | l l l @{}}
         $\overline{\rm MS}$& LO & NLO & NNLO  \\
         \midrule
         $a^d_0$ & ${8.20^{+4.20}_{-1.94}}$ & ${10.40^{+0.81}_{-1.45}}$ & $
         {11.40^{+0.55}_{-0.95}}$ \\[5pt]
         $b^d_0$ & ${0.069^{+0.037}_{-0.020}}$ & ${0.112^{+0.043}_{-0.020}}$ & ${0.134^{+0.042}_{-0.022}}$\\[5pt]
         $c^d_0$ & 
         ${-69.7^{+17.7}_{-35.0}}$ & $
         {-64.1^{+14.2}_{-0.0}}$ & $
         {-58.3^{+4.0}_{-1.9}}$ 
    \end{tabular}
    &
    \begin{tabular}{@{} l | l l l @{}}
         PS & LO & NLO & NNLO  \\
         \midrule
         $a^d_0$ & ${9.53^{+1.17}_{-0.39}}$ & ${11.11^{+0.22}_{-0.50}}$ & ${{ 11.71}^{+0.15}_{-0.35}}$ \\[5pt]
         $b^d_0$ & ${0.081^{+0.011}_{-0.009}}$ & ${0.122^{+0.027}_{-0.008}}$ & ${0.140^{+0.034}_{-0.015}}$ \\[5pt]
         $c^d_0$ & ${-81.0^{+5.2}_{-9.4}}$ & ${-66.2^{+8.7}_{-2.1}}$ & 
         ${-57.6^{+3.5}_{-3.3}}$
    \end{tabular}
    \end{tabular}
    \caption{Updated results for the values $a^d_0$, $b^d_0$ and $c^d_0$ in the $\overline{\textrm{MS}}$ (left) and PS (right) schemes for the $B_d$ system. The uncertainty shown here is the perturbative scale uncertainty obtained from varying $\mu_1=\mu_b=\mu_c$ simultaneously between $\SI{2.1}{GeV}$ and $\SI{8.4}{GeV}$.}
    \label{tab:afs_d}
\end{table}

\begin{table}[t]
    \centering
    \begin{tabular}{cc}
    \begin{tabular}{@{} l | l l l @{}}
         $\overline{\rm MS}$& LO & NLO & NNLO  \\
         \midrule
         $a^s_0$ & ${8.23^{+4.21}_{-1.94}}$ & ${10.43^{+0.82}_{-1.46}}$ & ${11.44^{+0.56}_{-0.96}}$ \\[5pt]
         $b^s_0$ & ${0.072^{+0.038}_{-0.021}}$ & ${0.115^{+0.043}_{-0.021}}$ & ${0.137^{+0.042}_{-0.022}}$\\[5pt]
         $c^s_0$ & ${-71.1^{+18.2}_{-35.8}}$ & ${-65.4^{+14.5}_{-0.0}}$ & ${-59.6^{+3.9}_{-1.9}}$ 
    \end{tabular}
    &
    \begin{tabular}{@{} l | l l l @{}}
         PS & LO & NLO & NNLO  \\
         \midrule
         $a^s_0$ & ${9.56^{+1.17}_{-0.39}}$ & ${11.15^{+0.23}_{-0.50}}$ & ${11.76^{+0.16}_{-0.35}}$ \\[5pt]
         $b^s_0$ & 
         ${{0.083}^{+0.011}_{-0.009}}$ & ${0.125^{+0.026}_{-0.007}}$ & ${0.143^{+0.034}_{-0.015}}$ \\[5pt]
         $c^s_0$ & 
         ${-82.6^{+5.4}_{-9.6}}$ & 
         ${-67.6^{+8.9}_{-2.1}}$ & 
         ${-58.9^{+3.5}_{-3.3}}$
    \end{tabular}
    \end{tabular}
    \caption{Updated results for the values $a^s_0$, $b^s_0$ and $c^s_0$ in the $\overline{\textrm{MS}}$ (left) and PS (right) schemes for the $B_s$ system. The uncertainty shown here is the perturbative scale uncertainty obtained from varying $\mu_1=\mu_b=\mu_c$ simultaneously between $\SI{2.1}{GeV}$ and $\SI{8.4}{GeV}$. }
    \label{tab:afs_s}
\end{table}

\begin{figure}[bt]
 \begin{center}
      \includegraphics[width=0.7\textwidth]{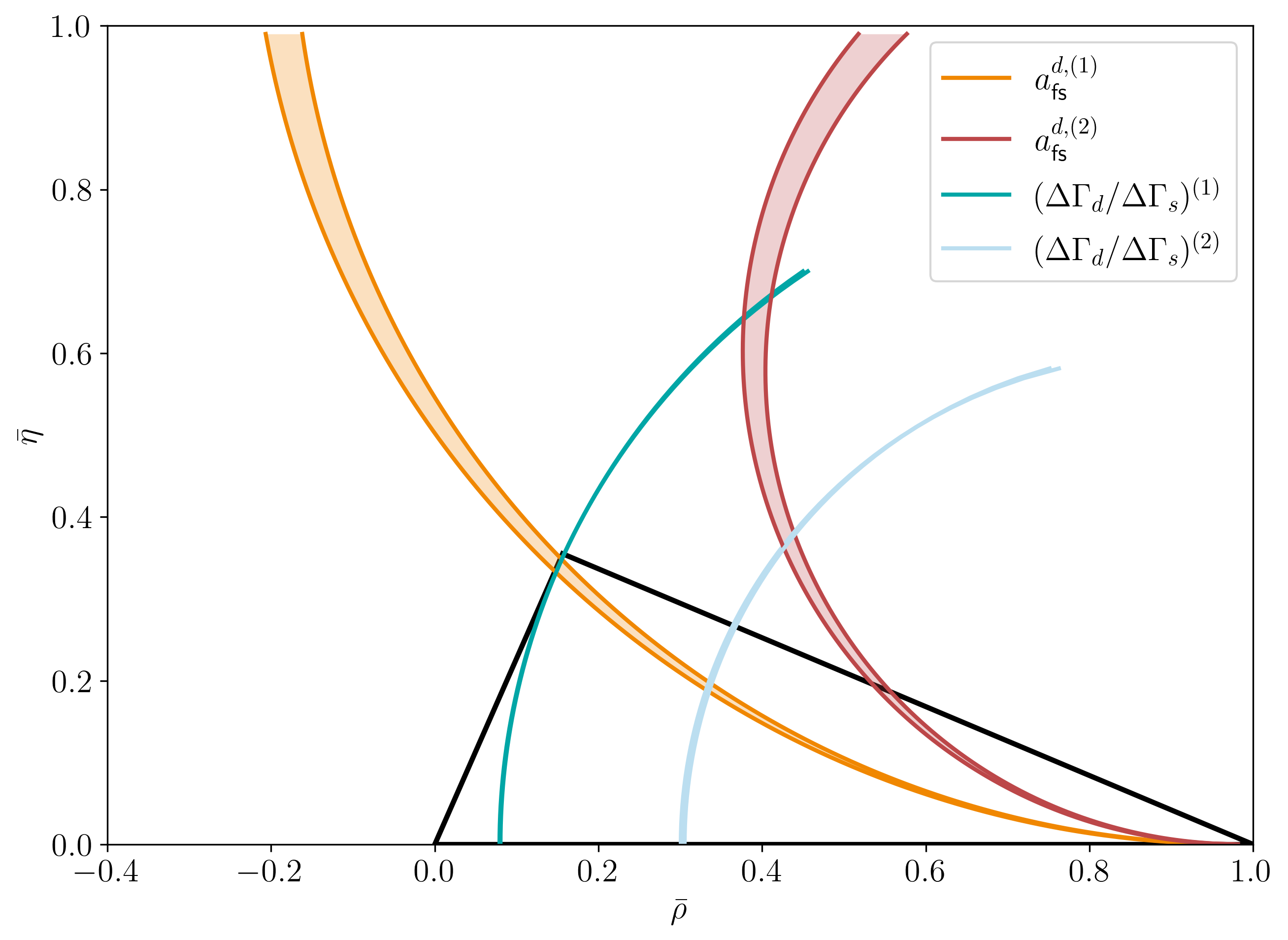}
  \end{center}    
\caption{\label{fig:double_constraints_CKM} 
      Constraints on the apex of the CKM triangle in the PS scheme. The bands correspond to the perturbative uncertainty of the leading-$1/m_b$ term, which have been added in quadrature, and are plotted for $a_\text{fs}^{d,(1)} =  -5 \times 10^{-4}$, $ a_\text{fs}^{d,(2)} =-1 \times 10^{-3}$, $(\Delta\Gamma_d/\Delta\Gamma_s)^{(1)}=0.029$ and $(\Delta\Gamma_d/\Delta\Gamma_s)^{(2)}=0.0145$.}    
\end{figure}

Using the numerical values of $a$, $b$ and $c$ given here, the constraints shown in Fig.~\ref{fig:double_constraints_CKM} can be plotted, where the error bands only include the uncertainty from the scale variation of the leading-$1/m_b$ term. To motivate more accurate determinations of the quantities $a_\text{fs}^d$ and $\Delta\Gamma_d$, we have shown
the constraints stemming from hypothetical measurements differing by a factor of two. Since all terms involving $a/c$ or $b/c$ are small, $\dg_d/\dg_s$ can be predicted with high precision. The factor $|\lambda_t^d|^2/|\lambda_t^s|^2$ is the same as in $\dm_d/\dm_s$, so that the curve in the $(\bar\rho,\bar\eta)$ plane looks very similar to the corresponding circle around $(1,0)$ with radius $R_t$ shown in light red in Fig.~\ref{fig:ut}. Since $\dg_d/\dg_s$ and $\dm_d/\dm_s$ depend on new-physics parameters in a very different way, this feature can be used to identify new physics in either ratio or in other quantities entering the global fit of $(\bar\rho,\bar\eta)$.

\begin{figure}[bt]
  \begin{center}
      \includegraphics[width=0.7\textwidth]{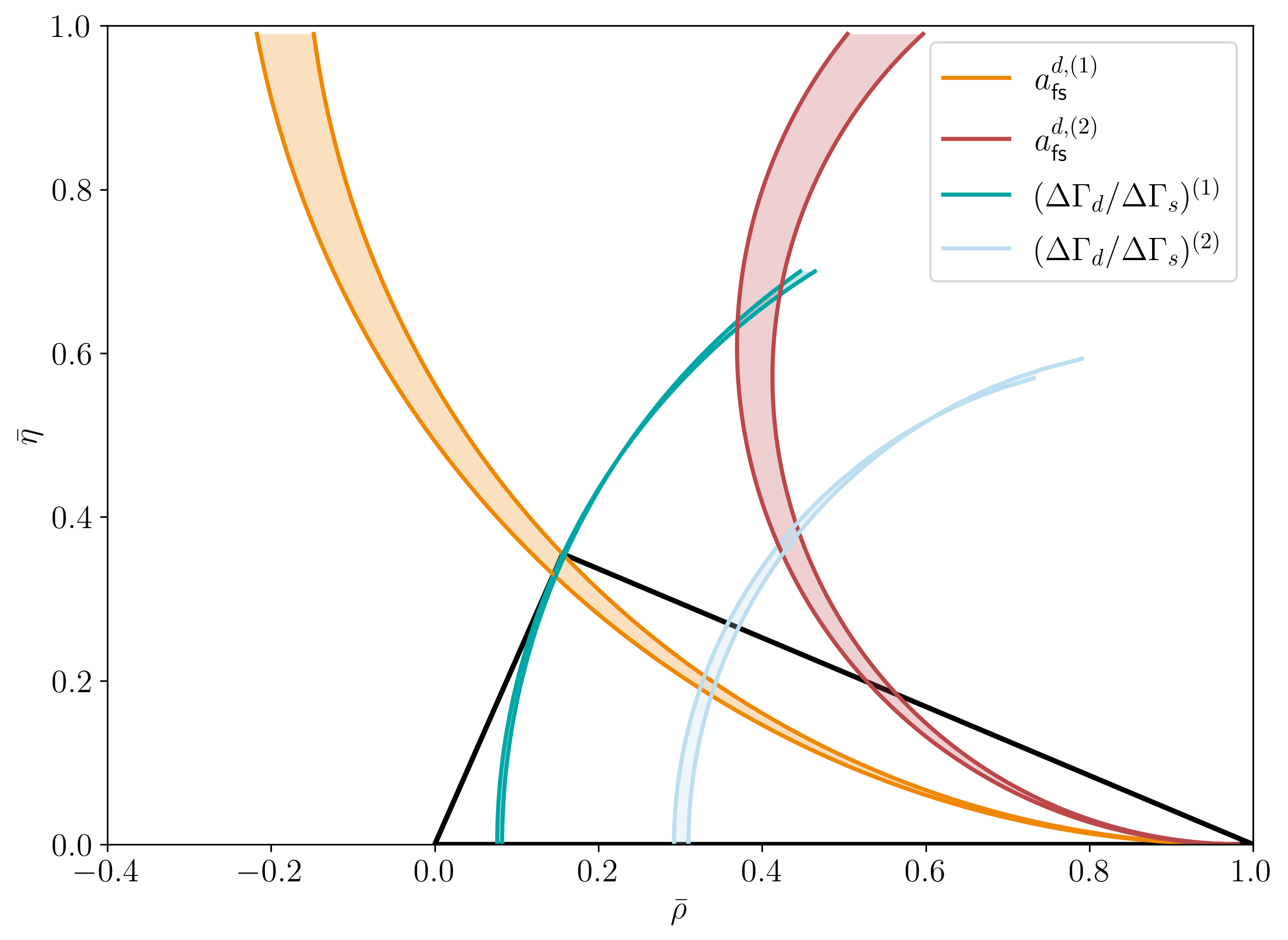}
  \end{center}    \caption{\label{fig:double_constraints_CKM_all} 
      Constraints on the apex of the CKM triangle in the PS scheme. The bands correspond to the full uncertainty and are plotted for $a_\text{fs}^{d,(1)} =  -5 \times 10^{-4}$, $ a_\text{fs}^{d,(2)} =-1 \times 10^{-3}$, $(\Delta\Gamma_d/\Delta\Gamma_s)^{(1)}=0.029$ and $(\Delta\Gamma_d/\Delta\Gamma_s)^{(2)}=0.0145$. The perturbative uncertainties of the leading-power and $1/m_b$-suppressed contributions have been added in quadrature here.}    
\end{figure}

To illustrate the size of the uncertainties of all input parameters, including the non-perturbative matrix elements, compared with the perturbative uncertainties of our calculations, we also show the constraints on the apex of the CKM triangle with all uncertainties combined in Fig.~\ref{fig:double_constraints_CKM_all}. We observe that the bands for $\Delta\Gamma_d/\Delta\Gamma_s$ are still very narrow and the current status of theoretical predictions lead to strong constraints on the apex of the unitarity triangle. This is possible because the ratio of the leading-$1/m_b$ matrix elements is proportional to $\xi^2$, which has been determined very accurately on the lattice.

\subsection{Numerical results independent of CKM values and hadronic matrix elements}

In this section, we present our results independent of any input values pertaining to the CKM matrix or the hadronic matrix elements, which carry most of the input uncertainties. Following Eq.~\eqref{eq:gabc}, we write
\begin{align}
 \frac{\Gamma_{12}^q}{M_{12}^q} &=\, 10^{-4} \times \sum_O  \left[ c_O + a_O \frac{\lambda_u^q}{\lambda_t^q} + b_O \frac{\lambda_u^{q\, 2}}{\lambda_t^{q\, 2}}\right]  \frac{\bra{B_q} O \ket{\bar{B}_q}}{\bra{B_q} Q \ket{\bar{B}_q}} \,,
 \label{eq:gabc_no_ome}
\end{align}
where the sum is over the set of operators
\begin{equation}
    O \in \{Q, \widetilde{Q}_S, R_0, R_1, \widetilde{R}_1, \widetilde{R}_2, \widetilde{R}_3 \}\,.
\end{equation}
Note that the matrix elements of $R_2$ and $R_3$ were eliminated using~\cite{Beneke:1996gn}
\begin{align}
    R_2 &= - \widetilde{R}_2\,, \nonumber\\
    R_3 &= \widetilde{R}_3 + \frac12 \widetilde{R}_2\,,
\end{align}
which holds up to corrections of higher order in $\alpha_s$ and $1/m_b$.
In the parametrisation of Eq.~\eqref{eq:gabc_no_ome}, all input parameters that depend on the specific $B_q$ mesons was factored out such that the constants $a_O$, $b_O$ and $c_O$ are the same for $B_s$ and $B_d$.
The results are given in Tabs.~\ref{tab:abc_general_MS}, \ref{tab:abc_general_PS} and \ref{tab:abc_general_mb}, where the $1/m_b$-suppressed terms do not differ between the two schemes since they are only know to LO. The uncertainties of all coefficients are dominated by the scale uncertainty since the remaining input parameters have a minor impact on the final result and are known relatively precisely.
\begin{table}[t]
    \centering
    \begin{tabular}{@{} l | l l l @{}}
         $\overline{\text{MS}}$& $c_O$ & $a_O$ & $b_O$ \\
         \midrule
         $Q$ & ${-42.6}^{+4.4}_{-1.8}\pm0.5$  & ${10.9}^{+0.46}_{-0.90}\pm0.21$ & ${0.108}^{+0.038}_{-0.023}\pm0.005$  \\[5pt]
         $\widetilde{Q}_S$ & ${-84.6}^{+0.88}_{-2.90}\pm0.99$ & ${2.66}^{+0.48}_{-0.28}\pm0.11$ & ${0.141}^{+0.023}_{-0.000}\pm0.011$ 
    \end{tabular}
    \caption{Results in the $\overline{\text{MS}}$ scheme for the individual matrix element coefficients with the two-sided perturbative and symmetrised input uncertainty. The scale uncertainty is obtained from varying $\mu_1=\mu_b=\mu_c$ simultaneously between $\SI{2.1}{GeV}$ and $\SI{8.4}{GeV}$.}
    \label{tab:abc_general_MS}
\end{table}
\begin{table}[t]
    \centering
    \begin{tabular}{@{} l | l l l @{}}
         PS & $c_O$ & $a_O$ & $b_O$ \\
         \midrule
         $Q$ & ${-41.9}^{+3.7}_{-2.7}\pm0.5$  & ${11.20}^{+0.08}_{-0.32}\pm0.23$ & ${0.115}^{+0.030}_{-0.017}\pm0.006$  \\[5pt]
         $\widetilde{Q}_S$ & ${-84.4}^{+0.8}_{-3.0}\pm1.0$ & ${2.74}^{+0.39}_{-0.16}\pm0.11$ & ${0.139}^{+0.022}_{-0.000}\pm0.012$ 
    \end{tabular}
    \caption{Results in the PS scheme for the individual matrix element coefficients with the two-sided perturbative and symmetrised input uncertainty. The scale uncertainty is obtained from varying $\mu_1=\mu_b=\mu_c$ simultaneously between $\SI{2.1}{GeV}$ and $\SI{8.4}{GeV}$.}
    \label{tab:abc_general_PS}
\end{table}
\begin{table}[t]
    \centering
    \begin{tabular}{@{} l | l l l @{}}
         $1/m_b$& $c_O$ & $a_O$ & $b_O$ \\
         \midrule
         $R_0$  & ${27.9}^{+4.3}_{-5.2}\pm0.5$ & ${-0.493}^{+0.093}_{-0.076}\pm0.019$ & ${-0.0449}^{+0.0084}_{-0.0069}\pm0.0024$ \\[5pt]
         $R_1$ & ${-55.8}^{+10.5}_{-8.6}\pm0.9$ & ${0.99}^{+0.15}_{-0.19}\pm0.04$ & ${0.090}^{+0.014}_{-0.017}\pm0.005$\\[5pt]
         $\widetilde{R}_1$ & ${172}^{+22}_{-14}\pm2$ & ${-3.05}^{+0.25}_{-0.39}\pm0.11$ & ${-0.277}^{+0.023}_{-0.036}\pm0.015$\\[5pt]
         $\widetilde{R}_2$ & ${231}^{+31}_{-25}\pm3$ & ${-0.73}^{+0.16}_{-0.12}\pm0.02$ & ${0.0615}^{+0.0058}_{-0.0000}\pm0.0036$\\[5pt]
         $\widetilde{R}_3$ & ${3.89}^{+0.46}_{-0.12}\pm0.14$ & ${4.47}^{+0.53}_{-0.14}\pm0.17$ & ${0.579}^{+0.068}_{-0.018}\pm0.032$
    \end{tabular}
    \caption{Results for the $1/m_b$-suppressed matrix element coefficients with the two-sided perturbative and symmetrised input uncertainty. The scale uncertainty is obtained from varying $\mu_1=\mu_b=\mu_c$ simultaneously between $\SI{2.1}{GeV}$ and $\SI{8.4}{GeV}$.}
    \label{tab:abc_general_mb}
\end{table}

The values given in Tabs.~\ref{tab:abc_general_MS}, \ref{tab:abc_general_PS} and \ref{tab:abc_general_mb} can be used to update, e.g.~the constraints on the CKM triangle as shown in Fig.~\ref{fig:double_constraints_CKM_all} once more accurate results for input values like the hadronic matrix elements become available. It is worth noting that the constraints from $\Delta\Gamma_d/\Delta\Gamma_s$ are essentially unchanged with the current input values whether the simplified formula from Eq.~\eqref{eq:drat} or the full parametrisation in Eq.~\eqref{eq:gabc_no_ome} is used.

%- }}}

%- {{{ Conclusions

\section{\label{sec::concl}Conclusions}
In this paper, we complete the NNLO corrections from leading-power operators
to the decay width difference and the charge-parity asymmetry.  We
include all current-current and penguin operators in the three-loop
contributions of the $|\Delta B|=1$ part and incorporate
the full charm quark mass dependence through deep expansions in
$m_c/m_b$. New ingredients from this paper are the penguin contributions
at NNLO and their full charm quark mass dependence at NLO.

We perform a detailed phenomenological analysis of $\Delta\Gamma_q$
and $a_{\rm fs}^q$ both for the $B_s$ and the $B_d$ system.  The
inclusion of the penguin contribution at NNLO leads to shifts
below {2\%} and { 0.5\%} for $\Delta \Gamma_q$ and $a_\text{fs}^q$ respectively for the central values of the renormalisation scale.
However, we observe a further stabilisation of the dependence on the
renormalisation scale which we use to estimate the uncertainties from
unknown higher-order corrections. They amount to about $\pm 7\%$ for
$\Delta\Gamma_q$ and $\pm 5\%$ for $a_{\rm fs}^q$.  A further
reduction requires a four-loop calculation
in the $|\Delta B|=1$ theory accompanied by a three-loop calculation
on the $|\Delta B|=2$ side together with a proper definition
of the corresponding evanescent operators. Such a calculation will
most likely not be available in the foreseeable future.

In our prediction the dominant uncertainty stems from $1/m_b$ corrections.
In order to allow for a simple incorporation of future improvements,
we provide ready-to-use formulae for $\Delta\Gamma_q/\Delta M_q$
where updated values for the leading and $1/m_b$-suppressed
bag parameters can be inserted in a straightforward way.

A phenomenological interesting quantity is the
double ratio $(\Delta\Gamma_s/\Delta M_s) / (\Delta\Gamma_d/\Delta M_d)$
since hadronic uncertainties cancel to a large extent, 
and the perturbative uncertainty is reduced to the sub-percent level. We calculate the most accurate prediction of $\Delta \Gamma_d$ to date with a combined uncertainty of around $6\%$, which can be further reduced to the sub-percent level with more accurate determinations of the hadronic bag parameters. 

Finally, we combine the predictions for $a_{\rm fs}^d$ and
$\Delta\Gamma_d / \Delta\Gamma_s$ to show how the apex of the CKM
triangle can be constrained from mixing observables in the $B_d$ and
$B_s$ system alone, without the information from a global fit to CKM
parameters. We find that the constraints from $a_{\rm fs}^d$ and in particular $\Delta\Gamma_d / \Delta\Gamma_s$ yield very stringent constraints given a particular set of measurements, which motivates more accurate experimental determinations of the $B$ mixing observables.

%- }}}

\section*{Acknowledgements}

The work was supported by the Deutsche Forschungsgemeinschaft (DFG, German
Research Foundation) under grant 396021762 --- TRR 257 ``Particle Physics
Phenomenology after the Higgs Discovery''.
Pascal Reeck would like to thank the Studienstiftung des deutschen Volkes for supporting him.

%\FloatBarrier

%- {{{ Bibl.:

\bibliographystyle{JHEP}
\bibliography{inspire_mod.bib,extra.bib}

@misc{progdata,
  author = "",
  date = "",
  howpublished = "Ancillary files at: \verb|https://www.ttp.kit.edu/preprints/2025/ttp25-054/|"
}

@article{Lange:2025fba,
    author = "Lange, Fabian and Usovitsch, Johann and Wu, Zihao",
    title = "{Kira 3: integral reduction with efficient seeding and optimized equation selection}",
    eprint = "2505.20197",
    archivePrefix = "arXiv",
    primaryClass = "hep-ph",
    reportNumber = "ZU-TH 39/25, HU-EP-25/17-RTG",
    month = "5",
    year = "2025"
}

@article{vanRitbergen:1998pn,
    author = "van Ritbergen, T. and Schellekens, A. N. and Vermaseren, J. A. M.",
    title = "{Group theory factors for Feynman diagrams}",
    eprint = "hep-ph/9802376",
    archivePrefix = "arXiv",
    reportNumber = "UM-TH-98-01, NIKHEF-98-004",
    doi = "10.1142/S0217751X99000038",
    journal = "Int. J. Mod. Phys. A",
    volume = "14",
    pages = "41--96",
    year = "1999"
}

@article{Ruijl:2017dtg,
    author = "Ruijl, Ben and Ueda, Takahiro and Vermaseren, Jos",
    title = "{FORM version 4.2}",
    eprint = "1707.06453",
    archivePrefix = "arXiv",
    primaryClass = "hep-ph",
    month = "7",
    year = "2017"
}

@article{Tentyukov:2007mu,
    author = "Tentyukov, M. and Vermaseren, J. A. M.",
    title = "{The Multithreaded version of FORM}",
    eprint = "hep-ph/0702279",
    archivePrefix = "arXiv",
    reportNumber = "NIKHEF-07-005, SFB-CPP-07-08, TTP07-06",
    doi = "10.1016/j.cpc.2010.04.009",
    journal = "Comput. Phys. Commun.",
    volume = "181",
    pages = "1419--1427",
    year = "2010"
}

@article{Vermaseren:2000nd,
    author = "Vermaseren, J. A. M.",
    title = "{New features of FORM}",
    eprint = "math-ph/0010025",
    archivePrefix = "arXiv",
    month = "10",
    year = "2000"
}

@article{Nogueira:1991ex,
    author = "Nogueira, Paulo",
    title = "{Automatic Feynman Graph Generation}",
    reportNumber = "IFM-7-91",
    doi = "10.1006/jcph.1993.1074",
    journal = "J. Comput. Phys.",
    volume = "105",
    pages = "279--289",
    year = "1993"
}

@article{HFLAV:2024ctg,
    author = "Banerjee, Swagato and others",
    collaboration = "Heavy Flavor Averaging Group (HFLAV)",
    title = "{Averages of $b$-hadron, $c$-hadron, and $\tau$-lepton properties as of 2023}",
    eprint = "2411.18639",
    archivePrefix = "arXiv",
    primaryClass = "hep-ex",
    month = "11",
    year = "2024"
}

@article{Chetyrkin:2017lif,
    author = "Chetyrkin, Konstantin G. and Kuhn, Johann H. and Maier, Andreas and Maierhofer, Philipp and Marquard, Peter and Steinhauser, Matthias and Sturm, Christian",
    title = "{Addendum to \textquotedblleft{}Charm and bottom quark masses: An update\textquotedblright{}}",
    eprint = "1710.04249",
    archivePrefix = "arXiv",
    primaryClass = "hep-ph",
    reportNumber = "DESY-17-152, FR-PHENO-2017-018, IPPP-17-72, TTP17-039",
    doi = "10.1103/PhysRevD.96.116007",
    month = "10",
    year = "2017",
    note = "[Addendum: Phys.Rev.D 96, 116007 (2017)]"
}

@article{Buras:1984pq,
    author = "Buras, A. J. and Slominski, W. and Steger, H.",
    title = "{$B^0 - \bar{B}^0$ Mixing, CP Violation and the B Meson Decay}",
    reportNumber = "MPI-PAE/PTh 7/84",
    doi = "10.1016/0550-3213(84)90437-1",
    journal = "Nucl. Phys. B",
    volume = "245",
    pages = "369--398",
    year = "1984"
}

@article{Gerlach:2021xtb,
    author = "Gerlach, Marvin and Nierste, Ulrich and Shtabovenko, Vladyslav and Steinhauser, Matthias",
    title = "{Two-loop QCD penguin contribution to the width difference in B$_{s}$\ensuremath{-}$ {\overline{B}}_s $ mixing}",
    eprint = "2106.05979",
    archivePrefix = "arXiv",
    primaryClass = "hep-ph",
    reportNumber = "P3H-21-040, TTP21-015",
    doi = "10.1007/JHEP07(2021)043",
    journal = "JHEP",
    volume = "07",
    pages = "043",
    year = "2021"
}

@article{ParticleDataGroup:2024cfk,
    author = "Navas, S. and others",
    collaboration = "Particle Data Group",
    title = "{Review of particle physics}",
    doi = "10.1103/PhysRevD.110.030001",
    journal = "Phys. Rev. D",
    volume = "110",
    number = "3",
    pages = "030001",
    year = "2024"
}

@article{Alonso-Alvarez:2021qfd,
    author = "Alonso-\'Alvarez, Gonzalo and Elor, Gilly and Escudero, Miguel",
    title = "{Collider signals of baryogenesis and dark matter from B mesons: A roadmap to discovery}",
    eprint = "2101.02706",
    archivePrefix = "arXiv",
    primaryClass = "hep-ph",
    reportNumber = "TUM-HEP 1299/20",
    doi = "10.1103/PhysRevD.104.035028",
    journal = "Phys. Rev. D",
    volume = "104",
    number = "3",
    pages = "035028",
    year = "2021"
}

@article{Maierhofer:2017gsa,
    author = {Maierh\"ofer, Philipp and Usovitsch, Johann and Uwer, Peter},
    title = "{Kira\textemdash{}A Feynman integral reduction program}",
    eprint = "1705.05610",
    archivePrefix = "arXiv",
    primaryClass = "hep-ph",
    doi = "10.1016/j.cpc.2018.04.012",
    journal = "Comput. Phys. Commun.",
    volume = "230",
    pages = "99--112",
    year = "2018"
}

@article{Beneke:2003az,
    author = "Beneke, Martin and Buchalla, Gerhard and Lenz, Alexander and Nierste, Ulrich",
    title = "{CP Asymmetry in Flavor Specific B Decays beyond Leading Logarithms}",
    eprint = "hep-ph/0307344",
    archivePrefix = "arXiv",
    reportNumber = "PITHA-03-05, LMU-16-03, FERMILAB-PUB-03-214-T",
    doi = "10.1016/j.physletb.2003.09.089",
    journal = "Phys. Lett. B",
    volume = "576",
    pages = "173--183",
    year = "2003"
}

@article{Fael:2022miw,
    author = {Fael, Matteo and Lange, Fabian and Sch\"onwald, Kay and Steinhauser, Matthias},
    title = "{Singlet and nonsinglet three-loop massive form factors}",
    eprint = "2207.00027",
    archivePrefix = "arXiv",
    primaryClass = "hep-ph",
    reportNumber = "TTP22-042, P3H-22-066",
    doi = "10.1103/PhysRevD.106.034029",
    journal = "Phys. Rev. D",
    volume = "106",
    number = "3",
    pages = "034029",
    year = "2022"
}

@inproceedings{Seidensticker:1999bb,
    author = "Seidensticker, T.",
    title = "{Automatic application of successive asymptotic expansions of Feynman diagrams}",
    booktitle = "{6th International Workshop on New Computing Techniques in Physics Research: Software Engineering, Artificial Intelligence Neural Nets, Genetic Algorithms, Symbolic Algebra, Automatic Calculation}",
    eprint = "hep-ph/9905298",
    archivePrefix = "arXiv",
    reportNumber = "TTP-99-22",
    month = "5",
    year = "1999"
}

@article{Gerlach:2022wgb,
    author = "Gerlach, Marvin and Nierste, Ulrich and Shtabovenko, Vladyslav and Steinhauser, Matthias",
    title = "{The width difference in $B - \bar B$ mixing at order $\alpha_s$ and beyond}",
    eprint = "2202.12305",
    archivePrefix = "arXiv",
    primaryClass = "hep-ph",
    reportNumber = "P3H-22-019, TTP22-012",
    doi = "10.1007/JHEP04(2022)006",
    journal = "JHEP",
    volume = "04",
    pages = "006",
    year = "2022"
}

@article{Davies:2019gnp,
    author = "Davies, Christine T. H. and Harrison, Judd and Lepage, G. Peter and Monahan, Christopher J. and Shigemitsu, Junko and Wingate, Matthew",
    collaboration = "HPQCD",
    title = "{Lattice QCD matrix elements for the ${B_s^0-\bar{B}_s^0}$ width difference beyond leading order}",
    eprint = "1910.00970",
    archivePrefix = "arXiv",
    primaryClass = "hep-lat",
    reportNumber = "INT-PUB-19-044, IPPP/19/77, JLAB-THY-19-3052",
    doi = "10.1103/PhysRevLett.124.082001",
    journal = "Phys. Rev. Lett.",
    volume = "124",
    number = "8",
    pages = "082001",
    year = "2020"
}

@article{Fael:2021kyg,
    author = {Fael, Matteo and Lange, Fabian and Sch\"onwald, Kay and Steinhauser, Matthias},
    title = "{A semi-analytic method to compute Feynman integrals applied to four-loop corrections to the $ \overline{\mathrm{MS}} $-pole quark mass relation}",
    eprint = "2106.05296",
    archivePrefix = "arXiv",
    primaryClass = "hep-ph",
    reportNumber = "P3H-21-041, TTP21-016",
    doi = "10.1007/JHEP09(2021)152",
    journal = "JHEP",
    volume = "09",
    pages = "152",
    year = "2021"
}

@article{Chetyrkin:1997gb,
    author = "Chetyrkin, Konstantin G. and Misiak, Mikolaj and Munz, Manfred",
    title = "{$|\Delta F| = 1$ nonleptonic effective Hamiltonian in a simpler scheme}",
    eprint = "hep-ph/9711280",
    archivePrefix = "arXiv",
    reportNumber = "MPI-PHT-97-51, TTP-97-44, ZU-TH-17-97, TUM-HEP-285-97, IFT-12-97",
    doi = "10.1016/S0550-3213(98)00131-X",
    journal = "Nucl. Phys. B",
    volume = "520",
    pages = "279--297",
    year = "1998"
}

@article{Harlander:1998cmq,
    author = "Harlander, R. and Seidensticker, T. and Steinhauser, M.",
    title = "{Complete corrections of $\mathcal{O}(\alpha \alpha_s)$ to the decay of the Z boson into bottom quarks}",
    eprint = "hep-ph/9712228",
    archivePrefix = "arXiv",
    reportNumber = "MPI-PHT-97-81, TTP-97-52",
    doi = "10.1016/S0370-2693(98)00220-2",
    journal = "Phys. Lett. B",
    volume = "426",
    pages = "125--132",
    year = "1998"
}

@article{Beneke:1996gn,
    author = "Beneke, M. and Buchalla, G. and Dunietz, I.",
    title = "{Width Difference in the $B_s-\bar{B_s}$ System}",
    eprint = "hep-ph/9605259",
    archivePrefix = "arXiv",
    reportNumber = "SLAC-PUB-7165, FERMILAB-PUB-96-095-T",
    doi = "10.1103/PhysRevD.54.4419",
    journal = "Phys. Rev. D",
    volume = "54",
    pages = "4419--4431",
    year = "1996",
    note = "[Erratum: Phys.Rev.D 83, 119902 (2011)]"
}

@article{Asatrian:2017qaz,
    author = "Asatrian, H. M. and Hovhannisyan, Artyom and Nierste, Ulrich and Yeghiazaryan, Arsen",
    title = "{Towards next-to-next-to-leading-log accuracy for the width difference in the $B_s-\bar{B}_s$ system: fermionic contributions to order $(m_c/m_b)^0$ and $(m_c/m_b)^1$}",
    eprint = "1709.02160",
    archivePrefix = "arXiv",
    primaryClass = "hep-ph",
    reportNumber = "TTP17-054",
    doi = "10.1007/JHEP10(2017)191",
    journal = "JHEP",
    volume = "10",
    pages = "191",
    year = "2017"
}

@article{Dowdall:2019bea,
    author = "Dowdall, R. J. and Davies, C. T. H. and Horgan, R. R. and Lepage, G. P. and Monahan, C. J. and Shigemitsu, J. and Wingate, M.",
    title = "{Neutral B-meson mixing from full lattice QCD at the physical point}",
    eprint = "1907.01025",
    archivePrefix = "arXiv",
    primaryClass = "hep-lat",
    reportNumber = "INT-PUB-19-031, JLAB-THY-19-3068",
    doi = "10.1103/PhysRevD.100.094508",
    journal = "Phys. Rev. D",
    volume = "100",
    number = "9",
    pages = "094508",
    year = "2019"
}

@article{Dowdall:2013tga,
    author = "Dowdall, R. J. and Davies, C. T. H. and Horgan, R. R. and Monahan, C. J. and Shigemitsu, J.",
    collaboration = "HPQCD",
    title = "{B-Meson Decay Constants from Improved Lattice Nonrelativistic QCD with Physical u, d, s, and c Quarks}",
    eprint = "1302.2644",
    archivePrefix = "arXiv",
    primaryClass = "hep-lat",
    doi = "10.1103/PhysRevLett.110.222003",
    journal = "Phys. Rev. Lett.",
    volume = "110",
    number = "22",
    pages = "222003",
    year = "2013"
}

@article{Hughes:2017spc,
    author = "Hughes, C. and Davies, C. T. H. and Monahan, C. J.",
    title = "{New methods for B meson decay constants and form factors from lattice NRQCD}",
    eprint = "1711.09981",
    archivePrefix = "arXiv",
    primaryClass = "hep-lat",
    reportNumber = "FERMILAB-PUB-17-547-T",
    doi = "10.1103/PhysRevD.97.054509",
    journal = "Phys. Rev. D",
    volume = "97",
    number = "5",
    pages = "054509",
    year = "2018"
}

@article{Gerlach:2022qnc,
    author = "Gerlach, Marvin and Herren, Florian and Lang, Martin",
    title = "{tapir: A tool for topologies, amplitudes, partial fraction decomposition and input for reductions}",
    eprint = "2201.05618",
    archivePrefix = "arXiv",
    primaryClass = "hep-ph",
    reportNumber = "TTP22-005, P3H-22-009, FERMILAB-PUB-22-025-T",
    doi = "10.1016/j.cpc.2022.108544",
    journal = "Comput. Phys. Commun.",
    volume = "282",
    pages = "108544",
    year = "2023"
}

@article{Bazavov:2017lyh,
    author = "Bazavov, A. and others",
    title = "{$B$- and $D$-meson leptonic decay constants from four-flavor lattice QCD}",
    eprint = "1712.09262",
    archivePrefix = "arXiv",
    primaryClass = "hep-lat",
    reportNumber = "FERMILAB-PUB-17/491-T, FERMILAB-PUB-17-491-T",
    doi = "10.1103/PhysRevD.98.074512",
    journal = "Phys. Rev. D",
    volume = "98",
    number = "7",
    pages = "074512",
    year = "2018"
}

@article{Hovhannisyan:2022miy,
    author = "Hovhannisyan, Artyom and Nierste, Ulrich",
    title = "{Addendum to: Towards next-to-next-to-leading-log accuracy for the width difference in the B$_{s}$\ensuremath{-}$ {\overline{B}}_s $ system: fermionic contributions to order (m$_{c}$/m$_{b}$)$^{0}$ and (m$_{c}$/m$_{b}$)$^{1}$}",
    eprint = "2204.11907",
    archivePrefix = "arXiv",
    primaryClass = "hep-ph",
    doi = "10.1007/JHEP06(2022)090",
    journal = "JHEP",
    volume = "06",
    pages = "090",
    year = "2022"
}

@article{Reeck:2024iwk,
    author = "Reeck, Pascal and Shtabovenko, Vladyslav and Steinhauser, Matthias",
    title = "{B meson mixing at NNLO: technical aspects}",
    eprint = "2405.14698",
    archivePrefix = "arXiv",
    primaryClass = "hep-ph",
    reportNumber = "P3H-24-031, TTP24-013, SI-HEP-2024-11",
    doi = "10.1007/JHEP08(2024)002",
    journal = "JHEP",
    volume = "08",
    pages = "002",
    year = "2024"
}

@article{Fael:2023zqr,
    author = {Fael, Matteo and Lange, Fabian and Sch\"onwald, Kay and Steinhauser, Matthias},
    title = "{Massive three-loop form factors: Anomaly contribution}",
    eprint = "2302.00693",
    archivePrefix = "arXiv",
    primaryClass = "hep-ph",
    reportNumber = "CERN-TH-2023-012, P3H-23-006, TTP23-002, ZU-TH 05/23",
    doi = "10.1103/PhysRevD.107.094017",
    journal = "Phys. Rev. D",
    volume = "107",
    number = "9",
    pages = "094017",
    year = "2023"
}

@article{Klappert:2020nbg,
    author = {Klappert, Jonas and Lange, Fabian and Maierh\"ofer, Philipp and Usovitsch, Johann},
    title = "{Integral reduction with Kira 2.0 and finite field methods}",
    eprint = "2008.06494",
    archivePrefix = "arXiv",
    primaryClass = "hep-ph",
    reportNumber = "TTK-20-24, P3H-20-041, FR-PHENO-2020-11, MITP/20-044",
    doi = "10.1016/j.cpc.2021.108024",
    journal = "Comput. Phys. Commun.",
    volume = "266",
    pages = "108024",
    year = "2021"
}

@article{LHCb:2021moh,
    author = "Aaij, R. and others",
    collaboration = "LHCb",
    title = "{Precise determination of the $B_{\mathrm{s}}^0$\textendash{}$\overline B_{\mathrm{s}}^0$ oscillation frequency}",
    eprint = "2104.04421",
    archivePrefix = "arXiv",
    primaryClass = "hep-ex",
    reportNumber = "LHCb-PAPER-2021-005, CERN-EP-2021-047",
    doi = "10.1038/s41567-021-01394-x",
    journal = "Nature Phys.",
    volume = "18",
    number = "1",
    pages = "1--5",
    year = "2022"
}

@article{Fael:2022rgm,
    author = {Fael, Matteo and Lange, Fabian and Sch\"onwald, Kay and Steinhauser, Matthias},
    title = "{Massive Vector Form Factors to Three Loops}",
    eprint = "2202.05276",
    archivePrefix = "arXiv",
    primaryClass = "hep-ph",
    reportNumber = "TTP22-009, P3H-22-016",
    doi = "10.1103/PhysRevLett.128.172003",
    journal = "Phys. Rev. Lett.",
    volume = "128",
    number = "17",
    pages = "172003",
    year = "2022"
}

@article{Gerlach:2022hoj,
    author = "Gerlach, Marvin and Nierste, Ulrich and Shtabovenko, Vladyslav and Steinhauser, Matthias",
    title = "{Width Difference in the $B$-$\bar B$ System at Next-to-Next-to-Leading Order of QCD}",
    eprint = "2205.07907",
    archivePrefix = "arXiv",
    primaryClass = "hep-ph",
    reportNumber = "TTP22-030, P3H-22-051",
    doi = "10.1103/PhysRevLett.129.102001",
    journal = "Phys. Rev. Lett.",
    volume = "129",
    number = "10",
    pages = "102001",
    year = "2022"
}

@article{Buras:1989xd,
    author = "Buras, Andrzej J. and Weisz, Peter H.",
    title = "{QCD Nonleading Corrections to Weak Decays in Dimensional Regularization and 't Hooft-Veltman Schemes}",
    reportNumber = "MPI-PAE/PTh-42/89, TUM-T32-189",
    doi = "10.1016/0550-3213(90)90223-Z",
    journal = "Nucl. Phys. B",
    volume = "333",
    pages = "66--99",
    year = "1990"
}

@article{ETM:2016nbo,
    author = "Bussone, A. and others",
    collaboration = "ETM",
    title = "{Mass of the b quark and B -meson decay constants from N$_f$=2+1+1 twisted-mass lattice QCD}",
    eprint = "1603.04306",
    archivePrefix = "arXiv",
    primaryClass = "hep-lat",
    doi = "10.1103/PhysRevD.93.114505",
    journal = "Phys. Rev. D",
    volume = "93",
    number = "11",
    pages = "114505",
    year = "2016"
}

@article{Kuipers:2012rf,
    author = "Kuipers, J. and Ueda, T. and Vermaseren, J. A. M. and Vollinga, J.",
    title = "{FORM version 4.0}",
    eprint = "1203.6543",
    archivePrefix = "arXiv",
    primaryClass = "cs.SC",
    reportNumber = "NIKHEF-2012-004, TTP12-008, SFB-CPP-12-15",
    doi = "10.1016/j.cpc.2012.12.028",
    journal = "Comput. Phys. Commun.",
    volume = "184",
    pages = "1453--1467",
    year = "2013"
}

@article{Lenz:2006hd,
    author = "Lenz, Alexander and Nierste, Ulrich",
    title = "{Theoretical update of $B_s - \bar{B}_s$ mixing}",
    eprint = "hep-ph/0612167",
    archivePrefix = "arXiv",
    reportNumber = "TTP06-31",
    doi = "10.1088/1126-6708/2007/06/072",
    journal = "JHEP",
    volume = "06",
    pages = "072",
    year = "2007"
}

@article{Gorbahn:2004my,
    author = "Gorbahn, Martin and Haisch, Ulrich",
    title = "{Effective Hamiltonian for non-leptonic $|\Delta F| = 1$ decays at NNLO in QCD}",
    eprint = "hep-ph/0411071",
    archivePrefix = "arXiv",
    reportNumber = "IPPP-04-66, DCPT-04-132, FERMILAB-PUB-04-281-T",
    doi = "10.1016/j.nuclphysb.2005.01.047",
    journal = "Nucl. Phys. B",
    volume = "713",
    pages = "291--332",
    year = "2005"
}

@article{Buras:1991jm,
    author = "Buras, Andrzej J. and Jamin, Matthias and Lautenbacher, M. E. and Weisz, Peter H.",
    title = "{Effective Hamiltonians for $\Delta S = 1$ and $\Delta B = 1$ nonleptonic decays beyond the leading logarithmic approximation}",
    reportNumber = "MPI-PAE-PTH-56-91, TUM-T31-16-91",
    doi = "10.1016/0550-3213(92)90345-C",
    journal = "Nucl. Phys. B",
    volume = "370",
    pages = "69--104",
    year = "1992",
    note = "[Addendum: Nucl.Phys.B 375, 501 (1992)]"
}

@article{Chetyrkin:2010ic,
    author = "Chetyrkin, K. and Kuhn, J. H. and Maier, A. and Maierhofer, P. and Marquard, P. and Steinhauser, M. and Sturm, C.",
    title = "{Precise Charm- and Bottom-Quark Masses: Theoretical and Experimental Uncertainties}",
    eprint = "1010.6157",
    archivePrefix = "arXiv",
    primaryClass = "hep-ph",
    reportNumber = "SFB-CPP-10-92, TTP10-45, MPP-2010-135, TTP10--45",
    doi = "10.1007/s11232-012-0024-7",
    journal = "Theor. Math. Phys.",
    volume = "170",
    pages = "217--228",
    year = "2012"
}

@article{Asatrian:2020zxa,
    author = "Asatrian, Hrachia M. and Asatryan, Hrachya H. and Hovhannisyan, Artyom and Nierste, Ulrich and Tumasyan, Sergey and Yeghiazaryan, Arsen",
    title = "{Penguin contribution to the width difference and $CP$ asymmetry in $B_q$-$\bar B_q$ mixing at order $\alpha_s^2 N_f$}",
    eprint = "2006.13227",
    archivePrefix = "arXiv",
    primaryClass = "hep-ph",
    reportNumber = "TTP-20-026, P3H-20-028",
    doi = "10.1103/PhysRevD.102.033007",
    journal = "Phys. Rev. D",
    volume = "102",
    number = "3",
    pages = "033007",
    year = "2020"
}

@article{Beneke:1998sy,
    author = "Beneke, M. and Buchalla, G. and Greub, C. and Lenz, A. and Nierste, U.",
    title = "{Next-to-leading order QCD corrections to the lifetime difference of $B_s$ mesons}",
    eprint = "hep-ph/9808385",
    archivePrefix = "arXiv",
    reportNumber = "CERN-TH-98-261, BUTP-98-20, MPI-PHT-98-58, DESY-98-101",
    doi = "10.1016/S0370-2693(99)00684-X",
    journal = "Phys. Lett. B",
    volume = "459",
    pages = "631--640",
    year = "1999"
}

@inproceedings{Nierste:2004uz,
    author = "Nierste, Ulrich",
    title = "{CP asymmetry in flavor-specific B decays}",
    booktitle = "{Proceedings of 39th Rencontres de Moriond on Electroweak Interactions and Unified Theories}",
    eprint = "hep-ph/0406300",
    archivePrefix = "arXiv",
    reportNumber = "FERMILAB-CONF-04-094-T",
    pages = "445--450",
    month = "6",
    year = "2004"
}

@article{Gerlach:2025tcx,
    author = "Gerlach, Marvin and Nierste, Ulrich and Reeck, Pascal and Shtabovenko, Vladyslav and Steinhauser, Matthias",
    title = "{Current-current operator contribution to the decay matrix in $B$-meson mixing at next-to-next-to-leading order of QCD}",
    eprint = "2505.22740",
    archivePrefix = "arXiv",
    primaryClass = "hep-ph",
    reportNumber = "P3H-25-034, SI-HEP-2025-11, TTP25-016",
    month = "5",
    year = "2025"
}

@article{Misiak:1994zw,
    author = "Misiak, Mikolaj and Munz, Manfred",
    title = "{Two loop mixing of dimension five flavor changing operators}",
    eprint = "hep-ph/9409454",
    archivePrefix = "arXiv",
    reportNumber = "TUM-T31-79-94",
    doi = "10.1016/0370-2693(94)01553-O",
    journal = "Phys. Lett. B",
    volume = "344",
    pages = "308--318",
    year = "1995"
}

@article{Khoze:1986fa,
    author = "Khoze, Valery A. and Shifman, Mikhail A. and Uraltsev, N. G. and Voloshin, M. B.",
    title = "{On Inclusive Hadronic Widths of Beautiful Particles}",
    reportNumber = "ITEP-86-156",
    journal = "Sov. J. Nucl. Phys.",
    volume = "46",
    pages = "112",
    year = "1987"
}

@article{Chetyrkin:1996vx,
    author = "Chetyrkin, Konstantin G. and Misiak, Mikolaj and Munz, Manfred",
    title = "{Weak radiative B meson decay beyond leading logarithms}",
    eprint = "hep-ph/9612313",
    archivePrefix = "arXiv",
    reportNumber = "ZU-TH-24-96, TUM-HEP-263-96, MPI-PHT-96-123",
    doi = "10.1016/S0370-2693(97)00324-9",
    journal = "Phys. Lett. B",
    volume = "400",
    pages = "206--219",
    year = "1997",
    note = "[Erratum: Phys.Lett.B 425, 414 (1998)]"
}

@article{Dighe:2001gc,
    author = "Dighe, A. S. and Hurth, T. and Kim, C. S. and Yoshikawa, T.",
    title = "{Measurement of the lifetime difference of $B_d$ mesons: Possible and worthwhile?}",
    eprint = "hep-ph/0109088",
    archivePrefix = "arXiv",
    reportNumber = "CERN-TH-2001-200, IFP-796-UNC, MPI-PHT-2001-27",
    doi = "10.1016/S0550-3213(01)00655-1",
    journal = "Nucl. Phys. B",
    volume = "624",
    pages = "377--404",
    year = "2002"
}

@article{Hovhannisyan:2025zev,
    author = "Hovhannisyan, Artyom and Nierste, Ulrich",
    title = "{Decay matrix of $B$-$\overline{B}$ mixing: Mixing of dimension-seven operators into dimension-six operators under renormalization}",
    eprint = "2511.17759",
    archivePrefix = "arXiv",
    primaryClass = "hep-ph",
    month = "11",
    year = "2025"
}

%- }}}

\end{document}